%% file: e672_psi97_prd_gg.tex
\documentstyle[preprint,eqsecnum,aps,epsf]{revtex}

\begin{document}
\draft
\title{
Production of $J/\psi$ mesons in $p  Be$ collisions 
at 530 and 800~GeV/$c$}

\input list_of_authors
\date{\today}
\maketitle

\input e672_symbols


\bigskip

\begin{center}
{\bf Abstract\\}
\end{center}

    We report on the production of \jpsi\ mesons in 530 and 
800~GeV/$c$ $p Be$ collisions in the Feynman-$x$ range 0.0$<$\xf $<$0.6. 
The \jpsi\ mesons were detected via decays into \mupmum. 
Differential distributions for \jpsi\ production have been measured as 
functions of \xf , \pts , and
cosine of the Gottfried-Jackson decay angle. These distributions
are compared 
with results on \jpsi\ production obtained in 515~GeV/$c$ 
$\pi^- Be$  collisions, measured by the same experiment, as well as 
with results 
from other experiments using incident protons. 

\vskip 5mm


\nopagebreak

\section{Introduction}

    Studies of charmonium production in hadron collisions provide important
information on both perturbative and non-perturbative QCD. Recent advances in
the understanding of quarkonium production have been stimulated by the
unexpectedly large cross sections observed for direct \jpsi\ and \psip\ production at
large \pt\  measured at the  Fermilab Tevatron~\cite{CDFD0}. These observations
have led to 
renewed interest in this field and to an improved theoretical understanding of
\jpsi\ production mechanisms. In particular,  a new model has been
developed to explain the CDF results: the color-octet mechanism~\cite{Braat,Gupta}. 
In the original color-singlet model~\cite{Baier},
charmonium  mesons retain the quantum numbers  of the  $c\bar{c}$ pairs
produced in the hard scatters, and thus each $J^{PC}$~state can only be directly
produced via the corresponding hard scattering color-singlet sub-processes. The
color-octet mechanism extends the color-singlet approach by taking into account
the production of $c\bar{c}$ pairs in a color-octet configuration accompanied
by a gluon. The color-octet state evolves into a color-singlet state via
emission of a soft gluon.

 Another recent analysis of \jpsi\ photo- and
hadro-production data~\cite {Halzen} 
resurrects the twenty-year-old color evaporation approach.
This approach differs from Ref.~\cite{Braat,Gupta} 
in the way the $c{\bar c}$ pair exchanges color with
the underlying event. It assumes multiple exchanges of soft gluons, whereas
such
interactions are suppressed in the color octet model by powers of $v$, the
relative velocity of the heavy quarks within the $c{\bar c}$ system. 
These multiple soft gluon
exchanges destroy the initial polarization of the heavy quark pair. Therefore,
polarization measurements may provide the best tool to distinguish between
these phenomenological models.

    In addition to the Tevatron collider data, a number of new results have become 
available
at fixed-target energies~\cite {Grib,Kore,Kowitt,Esen,Esso}. These measurements
provide further tests for the color octet and color evaporation
models~\cite{Gupta}.

    In this paper, we report on differential distributions for \jpsi\ 
production 
in  530~GeV/$c$ and  800~GeV/$c$ $p Be$ collisions, and compare these
distributions  with our
published results based upon  studies of 515~GeV/$c$ $\pi^- Be$ collisions~\cite{Grib}.
The pion and proton
data samples were collected during
the 1990 and 1991  Tevatron fixed-target runs, respectively, using  essentially the same 
apparatus.
In addition, during the 1991 fixed-target
run we recorded a lower statistics sample of data using a 515~GeV/$c$ negative pion beam to 
verify  the consistency of data samples from the two runs. 


    We also compare the 
distributions from the 800~GeV/$c$ incident proton data with the recently published 
results from
Fermilab experiments E789~\cite{Kowitt,Esen} and E771~\cite{Esso} obtained at the same
energy. The \xf\ range covered by this analysis (0.0$<$\xf $<$0.6)
complements 
the \xf\ ranges studied in Ref.~\cite{Kowitt} (0.3$<$\xf $<$0.95),
in Ref.~\cite{Esen} (-0.035$<$\xf $<$0.135), and in 
Ref.~\cite{Esso} (-0.05$<$\xf $<$0.25).

\section{Apparatus} 

    The experiment was performed in the Fermilab Meson West beam line using a
large-aperture, open-geometry spectrometer with the capability of 
triggering on the presence of
high-mass muon pairs~\cite{Grib,Blusk}.  The layout of the experiment 
is shown in Fig.~1.  A
brief description of the essential elements of the Meson West apparatus  as
implemented during the 1991 fixed-target run is presented below.

   The beam line included a 42-m-long differential Cerenkov counter capable of
tagging incident beam particles. The unseparated positive secondary  
beam at 530~GeV/$c$ was composed  primarily of protons with a small 
admixture ($<$3\%) of pions~\cite{Striley}. 
The 800~GeV/$c$ beam was a collimated primary proton beam extracted from the
Tevatron.  The beam intensity varied during the run, and at
its maximum was 2$\times 10^8$ protons per 23-second-long spill (at 57~second 
intervals). 

   A hadron shield consisting of 5~m of iron was located upstream of the target
to reduce background from off-axis hadrons and soft muons.  Two scintillator
veto walls were placed upstream of the hadron shield, and two others were
located downstream of the shield. These veto walls were used to identify high
energy muons that penetrated the iron.  An on-line veto in the dimuon trigger
rejected events containing coincidences between signals from either of the
upstream  veto walls and either of the downstream veto walls.

    The incident beam was defined via  a three plane ({\sl X, Y,} and 
{\sl U}) beam
hodoscope with millimeter-scale resolution,
 combined with a beam halo (BH) system of 
four scintillation counters
 located
upstream of the target. A beam particle was identified by 
at least  a two-fold coincidence from the
beam hodoscope in the absence of any signal from the BH system. The BH counters formed a 
rectangle 
with a 1~cm diameter 
hole centered on the beam. Signals from the beam
hodoscope were also employed to reject events containing more than one spatially
isolated incident beam particle. 


    The target consisted of two copper pieces, each 0.8~mm thick, followed by
a 15-cm-long liquid hydrogen target with 0.25~cm and 0.28~cm thick 
beryllium windows
upstream and downstream of the hydrogen target, respectively. A 2.54~cm thick
beryllium piece was located immediately downstream of the second window.
The total thickness of the target material corresponded 
to  $\approx $11\% of a proton interaction length.

    Three {\sl X-Y} modules  of 
silicon-strip detectors (SSDs) 
were located
upstream of the target to
measure the trajectories of incident beam particles.
Each module consisted of two single-sided planes.  Five more {\sl X-Y}
 SSD modules
were located downstream of the target. These modules were used to  measure
the trajectories of charged particles produced in the target and to reconstruct
primary and secondary vertices. The SSDs in the first downstream module had a
25-$\mu $m-pitch inner region and a 50-$\mu$m-pitch outer region; the SSDs in
the other modules had 50~$\mu $m~pitch throughout.

    A dipole magnet, producing a \pt~impulse of 0.45~GeV/$c$, was located
downstream of the SSDs. Four proportional wire chamber (PWC) modules were
located downstream of the dipole magnet. Each module contained four wire 
planes to
provide {\sl X, Y, U,} and {\sl V} measurements. (The {\sl U} and {\sl V}
 coordinates  were a pair of
orthogonal coordinates rotated by  $\deg{37}$ relative to the {\sl X} and {\sl 
Y}
coordinates.) The wires in each plane had  2.54~mm pitch. An area of from 6 to
26~cm$^2$ in the center of each plane had reduced sensitivity to provide
protection against beam particles. Two stations of straw tube
drift chambers, each with four planes in  the {\sl X} and four planes
in the {\sl Y} directions, were located adjacent to the outside PWC modules.  In the
analysis reported here, momentum measurements for charged tracks 
employed  all applicable tracking devices: the SSDs,
 the dipole magnet, the PWCs, and the straw tubes, 
yielding a momentum resolution of 
$\Delta p/p$~=~0.02\% $p$. In addition, the muon track candidates were constrained to
come from the primary vertex. (The straw tube data were not used in the results
reported in Ref.~\cite
{Grib}.)

    Interactions were detected by pairs of scintillation counters located
upstream and downstream of the dipole magnet. An interaction was
defined by signals from at least two of these four 
counters in coincidence with a
valid beam particle. The interaction rate was generally less than 1~MHz.

    A liquid-argon calorimeter (LAC)
was located downstream of the
magnetic spectrometer. The LAC contained both electromagnetic and hadronic
sections. The outer radius of the electromagnetic section was 165~cm; the inner
radius was 20~cm. A helium-gas-filled beam pipe was inserted along the axis  of
the LAC to minimize interactions of beam particles. The total LAC 
material
corresponded to more than 10~absorption lengths. 

    An iron and scintillator forward calorimeter was located downstream 
of the LAC to detect particles passing through  the beam pipe.
The forward calorimeter had an outer diameter of 1~m with a 3~cm diameter
axial beam
hole, and represented 10~absorption lengths of material.

    The muon detector
was located 20~m downstream of the
target and extended for 16~m.  The detector contained, in sequence, two muon 
PWC stations ($\mu$A, $\mu $B) with four planes each  ({\sl X, Y, U}, and {\sl 
V}), a beam
dump consisting of tungsten and steel imbedded in  concrete, an iron toroid
magnet producing an average \pt\ impulse of  1.3~GeV/$c$, and four more PWCs,
($\mu1 -\mu4$), each with three planes ({\sl X, U}, and {\sl V}).  
In these PWCs, the {\sl U} and {\sl V}
coordinates were at angles of $\deg{45}$ above and below the horizontal,
respectively. Iron, lead, and concrete shielding was interspersed between
chambers $\mu1$ through $\mu4$. Two muon hodoscope planes, H1 and H2
 (each with
sixteen  petal-shaped scintillation counters), were also located in this region.
The outer radius of the muon chambers and muon hodoscopes was 135~cm. 
The acceptance
of the muon spectrometer was limited by tapered axial holes, ranging
from 13~cm to 20~cm in radius, through the toroid magnet, the muon chambers, 
and the muon
hodoscopes. The hole in the toroid was filled with lead absorber.

    The muon detector elements H1, H2, and $\mu1$ through $\mu4$ were shielded
from hadrons by the material in the LAC, the forward calorimeter, 
the beam dump, 
the toroid, and the concrete shielding. Muons produced at the target 
with energies larger than 
$\approx$15~GeV 
penetrated this material, and all particles reaching the muon hodoscopes
were assumed to be muons.  Two or more hits in each of the muon hodoscopes were
required as part of the dimuon pretrigger;
the average hit
multiplicity in dimuon triggered events was 2.3 per hodoscope plane. 
In addition to the muon hodoscope 
requirement, the upstream veto wall requirements  completed
the pretrigger conditions. The pretrigger rate was 1.2 $\times 10^{-4}$
and 2 $\times 10^{-4}$ per live
interaction, for 530~GeV/$c$ and 800~GeV/$c$ incident protons, respectively.
 No radial 
dependence of the pretrigger counter efficiencies was observed.

    Events satisfying the dimuon pretrigger were evaluated 
by the dimuon trigger processor (DMTP),
which reconstructed space points in PWCs $\mu 1$ and $\mu 4$, formed muon
tracks (requiring an additional hit in either $\mu 2$ or $\mu 3$ along the
projected track trajectory), measured muon momenta from the estimated bend in
the toroid (assuming that the tracks originated in the target), and calculated
the dimuon invariant mass.  Trajectories of muon tracks reconstructed in the
downstream muon PWCs were projected to the center of the target. The
processor accepted only muon tracks with at least three hits  
within  roads around the projected
trajectory in both the
$\mu$A and $\mu$B chambers. A dimuon trigger resulted if any of the 
calculated dimuon masses
was above a preset threshold.
 A mass threshold of 1.0~GeV/$c^2$
resulted in a trigger rate of $1$ $\times 10^{-5}$ 
to $2 \times 10^{-5}$ per live interaction. 
 The average
DMTP processing time was 10 $\mu$s per pretrigger, which included 5 $\mu$s to
decode the muon chamber data. 
The \jpsi\ mass resolution of the DMTP was 550~MeV/$c^2$. 

    The toroid polarity was reversed several times during the run. In addition,
part of the data was recorded without the $\mu$A and $\mu$B hit requirement in
the trigger.

\section {Data}


    Table~I lists the numbers of dimuon triggers and approximate
integrated luminosities per
nucleon on $Be$
recorded during the 1991 fixed-target run for the different incident beams employed.
    For each dimuon-triggered event, 
the reconstructed muon tracks were
linked through the entire  detector.  
Only
events with at least two fully-linked muons were processed further. 
The remaining track
segments in the SSDs and upstream PWCs were used in the reconstruction
of other tracks 
and event vertices.  The distribution of the 
reconstructed primary 
vertices  along the
nominal beam direction $(Z)$, for events containing reconstructed dimuons in the
mass range 2.8~GeV/$c^2$ to 3.4~GeV/$c^2$, is shown in Fig.~2.

    Dimuons contributing to this analysis  came from events with primary
vertices in the beryllium targets, and had dimuon Feynman-$x$ (\xf\ =
2$p_{z}$/\sqs )
in the range 0.0$<$\xf $<$0.6 (0.0$<$ \xf $<$0.8 for $\pi^-$ data). 
Data from other targets 
are used to investigate nuclear effects.
Figure~3 shows the reconstructed
opposite-sign dimuon invariant  mass  distributions 
in the \jpsi\ mass region for the various data samples. 
The 
full width at half
maximum (FWHM) for the \jpsi\ signals 
varied from 120 to 135~MeV/$c^2$. Fits were performed
on the 
dimuon mass spectra in the mass range 1.5 to 5.0~GeV/$c^2$.  
These fits employed the sum of two exponentials for the continuum
background, and
resolution functions for the \jpsi\ and \psip\ resonances 
determined via Monte Carlo simulations.
The \jpsi\ masses resulting from the fits are within 1~MeV/$c^2$
of the Particle Data Group (PDG) value~\cite{PDG}. The \psip\ mass was fixed at the 
PDG value.
The numbers of
\jpsi s and \psip s obtained from these fits are listed in Table~I. 
 The
systematic uncertainties in the number of \jpsi\ and \psip\ combinations 
take into 
account variations  in the shapes  assumed for the signal and background
distributions. 

    To evaluate reconstruction efficiencies and acceptances, 
 $5\times 10^5$ Monte Carlo events containing \jpsi s were generated. 
The \xf\ and \pts\
distributions of the generated Monte Carlo \jpsi s were based on our 
previous
measurements~\cite{Grib,Abramov}.  We assumed that the 
\jpsi s decay isotropically
(consistent with observations to be described in the next section). 
These Monte Carlo events
also contained additional charged tracks with an average multiplicity 
consistent with
the data. The dimuons and associated particles were propagated through a GEANT
simulation of the Meson West spectrometer, which incorporated measured
spectrometer chamber efficiencies and instrumental noise  determined from 
our data.
The dimuon events were then reconstructed using 
the same tracking programs used
for reconstruction of the data.  We have evaluated the \jpsi\ reconstruction efficiency and
geometrical acceptance as a function of three variables: (i) $x_F$, the \jpsi\
Feynman-$x$;  (ii) \pts , the square of the \jpsi\ transverse momentum;  
and (iii) $cos \theta$,
the cosine of the Gottfried-Jackson decay angle between the $\mu ^{+}$ and the 
beam axis in the \jpsi\ rest frame. The products of acceptance and
reconstruction efficiency, $ a_{J/\psi \rightarrow \mu^+ \mu^-} \cdot
\varepsilon_{J/\psi \rightarrow \mu^+ \mu^-}$, were evaluated 
as one-dimensional distributions averaged over the other variable using an
iterative technique (see Fig.~4).  For a given  iteration,
the input
Monte Carlo \xf\ and  \pts\  distributions  were 
weighted to match those 
of the data corrected by the  acceptance and the reconstruction  efficiencies
determined in the previous iteration. The iterations were halted once
stability was achieved. In preparing Ref.~\cite{Grib} we demonstrated 
that this method of
evaluating acceptance and efficiency yielded cross section results consistent 
with a more sophisticated method based on a
two-dimensional acceptance surface over the \xf\ and \pts\ plane.
The resulting \xf\ and  \pts\ acceptances and efficiencies 
are almost independent
of the toroid polarity. 
But the product of acceptance  and efficiency as a function of \costh , integrated
over the \xf\ and \pts\ spectra,  shown in Fig.~4, exhibits a significant
dependence on the toroid polarity. This is a result of the difference in
acceptance as muons are either ``bent in'' or ``bent out'' by the toroid
magnet in the muon detector.

\section{\jpsi\ differential distributions}

    \jpsi\ candidates consisted of opposite sign dimuons with invariant
mass between 2.8~GeV/$c^2$ and 3.4~GeV/$c^2$ originating from the primary
vertex. Fits to the dimuon mass spectra outside the resonance band, 
similar to the ones described in
Section~III and shown in Fig.~3, for different regions of \xf ,  \pts , and
\costh\
indicate  significant variations in the background contribution as a function of
kinematic variables (for the 800~GeV/$c$ proton data).
 Therefore, in contrast to our previous analysis~\cite{Grib},
we have not carried out a two-constraint  kinematic fit 
for each \jpsi\ candidate
to improve the muon momentum  resolution.
Instead, we
performed a bin-by-bin background  subtraction to measure
the \jpsi\ differential distributions. The \costh\ distributions were
measured separately for the positive and negative toroid polarities, and
averaged as independent measurements. The effects of experimental
resolution were taken into account in the acceptance corrections.

    The \xf\ distributions for the incident proton data at 530~GeV/$c$ include 
corrections
for the 2.8\% incident $\pi^+$ contamination~\cite{Striley}. 
However, the Cerenkov
counter information was not available for the entire data set. In addition, 
the Cerenkov tag for protons that results in 99.6\% protons among the 
selected tagged
particles also results in $\approx$ 25\% reduction in the available luminosity.
Therefore, rather than use the beam Cerenkov 
counter
information on an event-by-event basis, we applied  iterative statistical
corrections to individual \xf\ bins. The corrections were based on the
measured \jpsi\ integrated cross sections per nucleon times \jpsi $\rightarrow
\mu^+\mu^-$ branching ratio for \xf $>$0 of 
    9.2 $\pm$ 2.0~nb~\cite{Abramov} and
12.9 $\pm$ 1.6~nb~\cite{Grib} for incident protons and pions, respectively. 
The shape of the \xf\ distribution for incident pions was taken
from Ref.~\cite {Grib}.
The size of the correction varied from 1\% at \xf =0.0
to 22\% at \xf = 0.55. For a given \xf\ bin, the systematic uncertainties due to this correction are
much smaller than the corresponding statistical uncertainties. 
The effects of beam contamination are negligible for the \pts , and \costh\ 
distributions,
which are similar for incident pions and protons.

    The \jpsi\ differential distributions  
 are
shown in Figs.~5 through 7, and tabulated in Tables~II and III. 
Data points were corrected
for the geometrical acceptances and reconstruction efficiencies  discussed in
the previous section. 
    The quoted uncertainties represent statistical and
systematic uncertainties added in quadrature. The systematic uncertainties 
include contributions due to uncertainties in the
background estimates, as well as pion contamination for the 530~GeV/$c$ 
incident proton data
sample. Additional systematic uncertainties, due to the trigger and off-line
reconstruction acceptance and 
efficiency calculations, were estimated from the variation of our results for
data subsamples taken under different running conditions. These contributions
are included
only
in the systematic uncertainties of the parameters of empirical fits employed to 
describe the data. 
Results of these
empirical fits are summarized in Tables~IV through VI.

    We used the parameterization: 
\begin{equation}
d\sigma/dx_F \propto {(1- {x_F})^c},
\end{equation}
 to describe the \jpsi\ \xf\ distribution 
for the proton-induced collisions [see 
Figs.~5(a) and ~6(a)]. The fit value of parameter $c$ increases from $c$ =
5.89 $\pm$ 0.14 ($stat$) $\pm$ 0.11 ($syst$) at 530~GeV/$c$ to $c$ = 6.18 
$\pm$ 0.12 ($stat$) $\pm$ 0.11 ($syst$) at 800~GeV/$c$.  
Our results are consistent with the general trend of increasing $c$ as the
center-of-mass energy increases (see Fig.~8). However, the values of parameter $c$ for
the incident beam momentum of 800~GeV/$c$, quoted by other experiments,
are 6.38$\pm$0.24 for a silicon target~\cite{Esso} and
4.91$\pm$0.18 for a gold target~\cite{Esen}.
(The silicon target data is restricted to the \xf\ range -0.05$<$\xf $<$0.25,
while the gold target data is restricted to -0.035$<$\xf $<$0.135).
In Fig.~9(a), we directly compare the \xf\ distributions from the three
experiments, normalized to the integral over the 0$<$\xf $<$0.135 range.
Our results agree very well with those of Ref.~\cite{Esso}, whereas the data
of Ref.~\cite{Esen} exhibit a harder \xf\ dependence. Quantatively, when a
fit of the \xf\ distribution for our 800~GeV/$c$ data is restricted
to the 0$<$\xf $<$0.15 range, that fit yields $c$ = 5.94 $\pm$ 0.43 with
a $\chi ^2$ per degree of freedom, $\chi ^2/ndf$ = 6.6/7.
Previous studies of the nuclear dependence of \jpsi\ production have
exhibited little nuclear dependence at small \xf~\cite{AZWaw}.
Comparisons of \jpsi\ \xf\ distributions
from our data using $H_2, Be,$ and $Cu$ targets, presented in Fig.~10,
do not show any significant dependence on the target nucleus.
A difference of one unit in the value of the parameter $c$ implies a relative
yield change between \xf = 0.0 and \xf = 0.25 by a factor of 0.75. The data show
that the ratios of the normalized distributions for different nuclear targets
are consistent with unity (within 25\%)  over the entire measured \xf\
range. Our observation of weak \xf\ dependence for nuclear effects
(within our large statistical uncertainties) is consistent with results
for $p Cu$ and $p Be$ collisions (\xf $>$0.3) reported in Ref.~\cite{Kowitt}.
The \xf\ distributions from Ref.~\cite{Kowitt} are compared
to our measurements in Fig.~9(b). Reasonable agreement is exhibited in the
overlapping \xf\ region.  We suspect that the discrepancy observed in the
small \xf\ region (discussed above) 
is due either to a large statistical fluctuation or
neglected experimental effects in some of the data sets.


    We have also used the parameterization proposed in Ref.~\cite{Likho}
to describe $d\sigma/dx_F$ :
\begin{equation}
d\sigma/dx_F \propto {(1-x_1 )}^\kappa {(1-x_2 )}^{\kappa}/(x_1
+x_2
),
\end{equation}
where
\begin{equation}  x_{1,2} = 0.5 (\sqrt{{x_F}^2 + 4\tau}\pm x_F),
\end{equation}
and $\tau$ = $M_{J/\psi}^2/s$. Here, $x_1$ represents the beam parton
fractional
momentum and $x_2$ the target parton fractional momentum. The
values for the parameter $\kappa$ obtained from the fit 
are 4.95 $\pm$ 0.17 ($stat$) $\pm$ 0.12 ($syst$) at 530~GeV/$c$ and  
4.95 $\pm$ 0.16 ($stat$) $\pm$ 0.12 ($syst$) at 800~GeV/$c$.
The corresponding  $\chi ^2/ndf$ are
24/13 and 30/13.
    A similar fit to the $\pi^-$ induced data at 515~GeV/$c$~\cite{Grib}, with
    the $ {(1-x_2 )}^{\kappa}$ factor replaced by $ {(1-x_2 )}^{\kappa + 2}$,
gave  $\kappa$=1.69 $\pm$ 0.04
with $\chi ^2/ndf$ = 50/33. 
    According to Ref.~\cite{Likho}, if gluon fusion is
the dominant mechanism responsible for \jpsi\ production via incident
protons, then the expected values of the parameter $\kappa$ are 5 for \jpsi s
produced predominantly through $\chi$ states decays, and 6 for direct \jpsi\
production. For incident pions, the gluon fusion mechanism results in $\kappa$
values of 3 and 4 for the indirect and direct \jpsi\ production, respectively. 
The
$q {\bar q}$ annihilation mechanism results in values of $\kappa$ smaller by one
unit. Our data indicate a larger difference between \jpsi\ production by
incident protons and pions than
predicted by simple counting rules.

    Gupta and Sridhar~\cite{Gupta} made a detailed comparison of the color-octet
model predictions with measured \xf\ distributions from several fixed-target
experiments, including our previous measurements described in 
Refs.~\cite{Grib,Abramov}. 
All values of the required non-perturbative matrix elements were
derived from other experiments, primarily from charmonium photoproduction 
data
and from \jpsi\ production data from the Tevatron.  In general, the
parameter-free predictions of the model for \xf -distributions agree with 
the proton and pion induced data both in magnitude and shape.
However, uncertainties in the model predictions due to scale uncertainties,
the various parton density sets, and non-perturbative matrix 
elements are still larger than the experimental uncertainties.

 
    The \pts\ spectra (Figs.~5b and 6b) were fit using the empirical form~\cite{Kaplan}

\begin{equation}   
    d\sigma/dp_T^2 \propto (1 + p_{T}^{2}/\beta^{2})^{-6},
\end{equation}
with parameters $\beta$ = [2.66 $\pm$ 0.04 ($stat$) $\pm$ 0.03 ($syst$)] GeV/$c$  and
[2.81 $\pm$ 0.03 ($stat$) $\pm$
0.03 ($syst$)]
GeV/$c$
for the 530~GeV/$c$ and 800~GeV/$c$ data, respectively.
The corresponding, \xf -integrated average
\jpsi\ transverse momenta, \avpt , are: [1.15 $\pm$
0.02 ($stat$)] GeV/$c$  and [1.22 $\pm$ 0.01 ($stat$)] GeV/$c$. 
As shown in Fig.~11,  the value of \avpt~grows 
slowly with center-of-mass energy. 
The comparison of \jpsi\ \pt\ distributions 
for the $H_2, Be,$
 and
$Cu$ targets, presented in Fig.~12, 
does not show any significant
dependence on
the target nucleus.  
 
    Fits to the \costh\ distribution in the range $\mid$\costh $\mid <$0.8
[Figs.~5(c) and 6(c)] using the functional form
 \begin{equation}  
    d\sigma/d(\cos \theta)\propto (1+\lambda\cos^{2}\theta),
\end{equation}
yield $\lambda$ = 0.01 $\pm$ 0.12 ($stat$) $\pm$ 0.09 ($syst$) 
 and $\lambda$ = -0.11 $\pm$ 0.12 ($stat$) $\pm$ 0.09 ($syst$) 
at 530~GeV/$c$ and 800~GeV/$c$, respectively. 
Figures~5(c) and
6(c) show the data for each toroid magnet
polarity separately, and the results  are similar.
As illustrated in Fig.~4, the acceptance and efficiency 
as a function of
\costh\ are sensitive to the toroid polarity. 
Nevertheless,
all data sets are consistent with unpolarized \jpsi\  production. 
The E771
Collaboration reported $\lambda$ = -0.09$\pm$0.12~\cite{Esso} in $p Si$
interactions at 800~GeV/$c$.
    Our measurement of $\lambda$
for \jpsi\ production from the high-statistic incident $\pi^-$ data sample 
gave $\lambda$ = -0.01$\pm$0.08~\cite{Grib}.  These results
confirm some of the earlier observations which reported small values of
$\lambda$~\cite{Akerlof}, and 
are consistent with expectations from the
color evaporation model
\cite{Halzen}.
These results for $\lambda$
are in disagreement with expectations from
the color octet model~\cite{Gupta},
which predicts a sizeable \jpsi\ transverse polarization, with mild energy
dependence (e.g.
0.31$<\lambda<0.63$ at \sqs = 15.3~GeV).

    The differential distributions for \jpsi\ production via 
incident 515~GeV/$c$ $\pi^-$ beams
are shown in Fig.~7. The new, lower statistics results, 
obtained during the same run as the proton data, are in a
good agreement with our published high-statistics results~\cite{Grib}.
The only substantial differences between the 
\jpsi\ differential distributions from the pion and proton induced data samples
are 
in the \xf\ distributions.
These differences are attributed to 
the harder distribution of gluons within the incident pion as compared 
to an incident proton. 

\section{Summary}

    We have studied the production of \jpsi\ mesons in the Feynman-$x$ range
0.0$<$\xf $<$0.6 
in 530~GeV/$c$ and 800~GeV/$c$ $p Be$ collisions.
We have parametrized the \jpsi\ differential distributions and
measured their dependence on center-of-mass energy. 
The \xf\ distributions of \jpsi\ production become more central as \sqs\
increases. For incident 800~GeV/$c$ protons, the \xf\ distribution observed in
our data ($Be$ target) is similar to the distribution reported by E771 ($Si$
target)\cite{Esso}, and differs from the distribution reported by E789 ($Au$
target)\cite{Esen}. Based upon studies of data from $H_2, Be,$ and $Cu$ targets, we are
inclined to believe that this difference is not likely to be explained by
nuclear effects.
The \xf\ distribution for 
\jpsi\ production in $\pi^- Be$ interactions is less central than the
corresponding distribution for $pB$e interactions at a similar beam energy.
    The data are consistent with unpolarized \jpsi\  production as
anticipated in the color evaporation model, but are in disagreement with
expectations from
the color octet model.

    We thank the staffs of all participating institutions, especially those
of Fermilab and the Institute for High Energy Physics (IHEP) at Protvino. This
work was supported by the U.~S. Department of Energy, the National Science
Foundation, and the Russian Ministries of Science and Atomic Energy.
	
\newpage

\newpage
\begin{center}
Table~I. Data sets and number of \jpsi\ and \psip\ combinations from
interactions on the $Be$
targets during the 1991 fixed-target run.
\end{center}
\begin{tabular}{|c|c|c|c|c|c|} \cline{1-6}
{ Beam  } & { Momentum } & { Integrated } & { Dimuon }& 
{$N_{J/\psi}$ } & { $N_{\psi(2S)}$}\\ 
&  &Luminosity & triggers & &\\ 
 & (GeV/$c$) &(pb$^{-1}$) & $10^3$ & &\\ \hline
$\pi^-$ & 515 & 1.4 &  902 &  806$\pm  32 \pm 13$ &  18 $\pm  10 \pm  6$ \\
$p$       & 530 & 6.4 & 2434 & 3607$\pm  65 \pm 28$ & 100 $\pm  18 \pm 15$ \\
$p$       & 800 & 7.3 & 6400 & 6540$\pm  90 \pm 43$ & 214 $\pm  27 \pm 20$ \\
\hline
\end{tabular}

\newpage
 \begin{center}
 Table~II.   Differential distributions for \jpsi\ production in
  530~GeV/$c$ $pBe$
 interactions.   The quoted uncertainties represent statistical and
systematic uncertainties added in quadrature. The systematic uncertainties
include contributions due to uncertainties in the
background estimates and  pion contamination.
 \end{center}
 \begin{tabular}{|lc|lc|lc|}     \hline
 ~~$x_F$ &  $  (1/\sigma)d\sigma/dx_F$   &
 ~~$p_T^2$  &  $  (1/\sigma)d\sigma/dp_T^2$~~&
 ~~$cos \theta$ & $(1/\sigma)d\sigma/dcos \theta$~ \\
    &   &
 ~(GeV/c)$^2$   &   1/(GeV/$c$)$^2$ &  & \\  \hline

  $ ~ 0.00$ -- $ 0.05 $ & $  5.7 \pm  1.6 $ &   $ 0.00$ -- $ 0.25 $ & $0.659 \pm0.028 $ &   $ -0.8$ -- $ -0.6 $ & $0.532 \pm0.038 $ \\
  $ ~ 0.05$ -- $ 0.10 $ & $ 4.56 \pm 0.34 $ &   $ 0.25$ -- $ 0.50 $ & $0.556 \pm0.026 $ &   $ -0.6$ -- $ -0.5 $ & $0.490 \pm0.027 $ \\
  $0.100$ -- $0.125 $ & $ 3.60 \pm 0.25 $ &   $ 0.50$ -- $ 0.75 $ & $0.401 \pm0.023 $ &   $ -0.5$ -- $ -0.4 $ & $0.523 \pm0.024 $ \\
  $0.125$ -- $0.150 $ & $ 2.90 \pm 0.19 $ &   $ 0.75$ -- $ 1.00 $ & $0.311 \pm0.021 $ &   $ -0.4$ -- $ -0.3 $ & $0.463 \pm0.020 $ \\
  $0.150$ -- $0.175 $ & $ 2.43 \pm 0.15 $ &   $ 1.00$ -- $ 1.25 $ & $0.305 \pm0.021 $ &   $ -0.3$ -- $ -0.2 $ & $0.507 \pm0.018 $ \\
  $0.175$ -- $0.200 $ & $ 1.92 \pm 0.12 $ &   $ 1.25$ -- $ 1.50 $ & $0.250 \pm0.019 $ &   $ -0.2$ -- $ -0.1 $ & $0.522 \pm0.020 $ \\
  $0.200$ -- $0.225 $ & $ 1.93 \pm 0.11 $ &   $ ~  1.5$ -- $  2.0 $ & $0.187 \pm0.012 $ &   $ -0.1$ -- $  0.0 $ & $0.502 \pm0.021 $ \\
  $0.225$ -- $0.250 $ & $1.183 \pm0.086 $ &   $ ~  2.0$ -- $  2.5 $ & $0.125 \pm0.010 $ &   $ ~~~  0.0$ -- $  0.1 $ & $0.539 \pm0.022 $ \\
  $0.250$ -- $0.275 $ & $1.185 \pm0.083 $ &   $ ~  2.5$ -- $  3.0 $ & $0.105 \pm0.009 $ &   $ ~~~  0.1$ -- $  0.2 $ & $0.485 \pm0.024 $ \\
  $0.275$ -- $0.300 $ & $0.860 \pm0.071 $ &   $ ~  3.0$ -- $  4.0 $ & $0.064 \pm0.005 $ &   $ ~~~  0.2$ -- $  0.3 $ & $0.480 \pm0.027 $ \\
  $ ~ 0.30$ -- $ 0.35 $ & $0.721 \pm0.045 $ &   $ ~  4.0$ -- $  5.0 $ & $0.053 \pm0.005 $ &   $ ~~~  0.3$ -- $  0.4 $ & $0.523 \pm0.032 $ \\
  $ ~ 0.35$ -- $ 0.40 $ & $0.452 \pm0.036 $ &   $ ~  5.0$ -- $  7.0 $ & $0.018 \pm0.002 $ &   $ ~~~  0.4$ -- $  0.5 $ & $0.423 \pm0.038 $ \\
  $ ~ 0.40$ -- $ 0.45 $ & $0.234 \pm0.027 $ &   $ ~  7.0$ -- $ 10.0 $ & $0.006 \pm0.001 $ &   $ ~~~  0.5$ -- $  0.6 $ & $0.543 \pm0.040 $ \\
  $ ~ 0.45$ -- $ 0.50 $ & $0.167 \pm0.023 $ &   &  &   $ ~~~  0.6$ -- $  0.8 $ & $ 0.96 \pm0.214 $ \\
  $ ~ 0.50$ -- $ 0.60 $ & $0.095 \pm0.014 $ &   &  &   & \\ \hline
\end{tabular}
\newpage
 \begin{center}
 Table~III.  Differential distributions for \jpsi\ production in
  800~GeV/$c$ $pBe$
 interactions.   The quoted uncertainties represent statistical uncertainties
and
systematic uncertainties in the
background estimates added in quadrature. 

 \end{center}
 \begin{tabular}{|lc|lc|lc|}     \hline
 ~~$x_F$ &  $  (1/\sigma)d\sigma/dx_F$   &
 ~~$p_T^2$  &  $  (1/\sigma)d\sigma/dp_T^2$~~&
 ~~$cos \theta$ & $(1/\sigma)d\sigma/dcos \theta$~ \\
    &   &
 ~(GeV/$c$)$^2$   &   1/(GeV/$c$)$^2 $ &  & \\  \hline

  $ ~ 0.00$ -- $ 0.05 $ & $ 6.14 \pm 0.33 $ &   $ 0.00$ -- $ 0.25 $ & $0.648 \pm0.021 $ &   $ -0.8$ -- $ -0.6 $ & $0.444 \pm0.038 $ \\
  $ ~ 0.05$ -- $ 0.08 $ & $ 4.79 \pm 0.22 $ &   $ 0.25$ -- $ 0.50 $ & $0.481 \pm0.019 $ &   $ -0.6$ -- $ -0.5 $ & $0.488 \pm0.030 $ \\
  $0.075$ -- $0.100 $ & $ 4.07 \pm 0.17 $ &   $ 0.50$ -- $ 0.75 $ & $0.369 \pm0.017 $ &   $ -0.5$ -- $ -0.4 $ & $0.503 \pm0.026 $ \\
  $0.100$ -- $0.125 $ & $ 3.56 \pm 0.14 $ &   $ 0.75$ -- $ 1.00 $ & $0.337 \pm0.016 $ &   $ -0.4$ -- $ -0.3 $ & $0.453 \pm0.024 $ \\
  $0.125$ -- $0.150 $ & $ 2.87 \pm 0.12 $ &   $ 1.00$ -- $ 1.25 $ & $0.289 \pm0.015 $ &   $ -0.3$ -- $ -0.2 $ & $0.476 \pm0.022 $ \\
  $0.150$ -- $0.175 $ & $ 2.29 \pm 0.10 $ &   $ 1.25$ -- $ 1.50 $ & $0.226 \pm0.013 $ &   $ -0.2$ -- $ -0.1 $ & $0.500 \pm0.022 $ \\
  $0.175$ -- $0.200 $ & $1.962 \pm0.094 $ &   $ ~  1.5$ -- $  2.0 $ & $0.180 \pm0.009 $ &   $ -0.1$ -- $  0.0 $ & $0.499 \pm0.023 $ \\
  $0.200$ -- $0.225 $ & $1.710 \pm0.085 $ &   $ ~  2.0$ -- $  2.5 $ & $0.144 \pm0.008 $ &   $ ~~~  0.0$ -- $  0.1 $ & $0.535 \pm0.023 $ \\
  $0.225$ -- $0.250 $ & $1.579 \pm0.076 $ &   $ ~  2.5$ -- $  3.0 $ & $0.111 \pm0.007 $ &   $ ~~~  0.1$ -- $  0.2 $ & $0.501 \pm0.024 $ \\
  $0.250$ -- $0.275 $ & $1.062 \pm0.071 $ &   $ ~  3.0$ -- $  4.0 $ & $0.069 \pm0.004 $ &   $ ~~~  0.2$ -- $  0.3 $ & $0.493 \pm0.027 $ \\
  $0.275$ -- $ 0.30 $ & $0.837 \pm0.067 $ &   $ ~  4.0$ -- $  5.0 $ & $0.054 \pm0.003 $ &   $ ~~~  0.3$ -- $  0.4 $ & $0.562 \pm0.033 $ \\
  $ ~ 0.30$ -- $ 0.35 $ & $0.594 \pm0.043 $ &   $ ~  5.0$ -- $  6.0 $ & $0.032 \pm0.003 $ &   $ ~~~  0.4$ -- $  0.5 $ & $0.468 \pm0.039 $ \\
  $ ~ 0.35$ -- $ 0.40 $ & $0.341 \pm0.037 $ &   $ ~  6.0$ -- $  8.0 $ & $0.013 \pm0.001 $ &   $ ~~~  0.5$ -- $  0.6 $ & $0.553 \pm0.054 $ \\
  $ ~ 0.40$ -- $ 0.45 $ & $0.305 \pm0.036 $ &   $ ~  8.0$ -- $ 10.0 $ & $0.007 \pm0.001 $ &   $ ~~~  0.6$ -- $  0.8 $ & $ 0.74 \pm0.116 $ \\
  $ ~ 0.45$ -- $ 0.55 $ & $0.128 \pm0.026 $ &   &  &   & \\ \hline
\end{tabular}
\newpage 

\begin{center}
Table~IV. Results of 
fits to the  $d\sigma/dx_F$ distributions for $J/\psi$ production.
\end{center}
\bigskip

\begin{tabular}{|c|c|c|c|c|c|} \hline 
 Beam               &  Momentum (GeV/$c$) & $c$ & $\chi ^2/ndf$ & $\kappa$ &
 $\chi ^2/ndf$\\ \hline
  $\pi^-$           &  515 [6]          & & &$1.69\pm0.04      $ & $50/33$ \\
   $p$                &  530              & $5.89\pm0.14\pm0.11$ & $21/13$ & 
4.95$\pm$0.17 $\pm$0.12&24/13\\
   $p$                &  800 & $6.18\pm0.12\pm0.11$ & $21/13$ & 
4.95$\pm$0.16$\pm$0.12& 30/13\\    \hline 
\end{tabular}

\bigskip 
\bigskip 

\begin{center}
Table~V. 
Results of fits to the  $d\sigma/dp_T^2$ distributions for $J/\psi$ production.
\end{center}

\bigskip
\begin{tabular}{|c|c|c|c|c|} \hline \hline                                             
 Beam               &  Momentum (GeV/$c$) & $\beta$   & $\chi ^2/ndf$  & 
$\langle p_T \rangle$ (GeV/$c$)   \\ \hline \hline
  $ \pi^-$ [6]      &  515  &                       &          & $1.17\pm0.02$    \\
   $p$                &  530  & $2.66\pm0.037\pm0.03$ & $19/11$  & $1.15\pm0.02$    \\
   $p$                &  800  & $2.81\pm0.029\pm0.03$ & $35/12$  
& $1.22\pm0.01$    \\  \hline 
\end{tabular}

\bigskip
\bigskip
\begin{center}
Table~VI. Results of fits to the $d\sigma/d(cos\theta)$ distributions for
$J/\psi$ production.
\end{center}

\bigskip
\begin{tabular}{|c|c|c|c|} \hline \hline                                             
 Beam               &  Momentum (GeV/$c$) & $\lambda$           & $\chi ^2/ndf$ \\ \hline
 $\pi^-$ [6]        &  515              & $-0.01\pm0.08$       & $33/34$         \\
   $p$                &  530              & $0.008\pm0.12\pm0.09$& $20/12$         \\
   $p$ &  800 & $-0.11\pm0.12\pm0.09$ & $19/12$ \\  \hline 
\end{tabular}

\newpage

\begin{center}  FIGURE CAPTIONS.
\end{center}

\bigskip

\begin{list}{}{}

\itemsep 0.25in

\item Figure~1.  Plan view of the Fermilab Meson West spectrometer as
configured for the 
1991 fixed-target run.

\item Figure~2.  Primary vertex Z-coordinate distribution for events containing
dimuons with reconstructed masses in the \jpsi\ mass range. Unmarked spikes in
the distribution upstream of the $Cu$ target and downstream of the $Be$ target
are due to interactions in the silicon-strip detectors.

\item  Figure~3.   The invariant mass distributions of $\mu^+\mu^-$ pairs 
for: (a) 530~GeV/$c$ incident protons, (b) 800~GeV/$c$ incident protons,
and (c) 515~GeV/$c$ incident $\pi^-$. The solid curve in each plot is a fit to 
the data; the dashed curve shows the
background contribution.

\item Figure~4.  Product of geometrical acceptance and  reconstruction 
efficiency for detection of \psimumu\
as a function of: (a,b) \xf , (c,d) \pts , and (e,f) $cos \theta$ 
for 530~GeV/$c$
(left column) and 800~GeV/$c$ (right column) incident protons.
The solid and dashed curves represent the results for the positive and 
negative toroid
polarity, respectively.

\item Figure~5.  
Differential distributions for \jpsi\ production 
as functions of: (a) \xf  , (b)
\pts\  ((GeV/$c)^{2}$), and  (c) $cos \theta$  for $p Be$ interactions at 
530~GeV/$c$. The $cos \theta$ distributions are shown separately for each
toroid magnet
polarity. 
The integrals of the distributions in (a) and (b) are normalized to unity.
The integrals of the $cos \theta$ distributions within the
-0.6$<cos \theta <0.6 $ range are normalized to 0.6.
The solid and dashed curves in (a) represent empirical fits to the data using
Eqs.~(4.1) and (4.2), respectively. 
The solid curves in (b) and (c) represent fits using the
functions shown in Eqs.~(4.4) and (4.5), respectively. 

\item Figure~6.  
Differential distributions for \jpsi\ production as functions of: (a) \xf  , (b)
\pts\  ((GeV/$c)^{2}$), and  (c) $cos \theta$  for $p Be$ interactions at 
800~GeV/$c$. The $cos \theta$ distributions are shown separately for each
toroid magnet
polarity. 
The integrals of the distributions in (a) and (b) are normalized to unity.
The integrals of the $cos \theta$ distributions within the 
-0.6$<cos \theta <$0.6 range 
are normalized to 0.6. 
The solid and dashed curves in (a) represent empirical fits to the data using
Eqs.~(4.1) and (4.2), respectively.
The solid curves in (b) and (c) represent fits using the
functions shown in Eqs.~(4.4) and (4.5), respectively.


\item Figure~7.  
Differential distributions for $J/\psi$ production 
as functions of: (a) \xf  , (b)
\pts\  ((GeV/$c)^{2}$), and  (c) $cos \theta$  for $\pi ^- Be$ interactions at 
515~GeV/$c$. 
The integrals of the distributions in (a) and (b) are normalized to unity.
The integral of the $cos \theta$ distribution within the
-0.6$< cos \theta  <$ 0.6 range
is normalized to 0.6.
The data of Ref.~\cite{Grib} are shown for comparison. 

\item Figure~8.  Center-of-mass energy dependence of the exponent $c$ in  the
parametrization  of the  \xf\ distributions for \jpsi\ production in
nucleon-nucleus
interactions. The \xf\  distributions for Refs.~[20,21,22] were converted into
the d$\sigma$/d\xf\ form from the quoted Lorentz invariant cross sections
$E d\sigma$/d\xf\ using the average value of the \jpsi\ transverse
momentum at a given center-of-mass energy.

\item Figure~9.  Differential distribution for $J/\psi$ production in 
800~GeV/$c$ $p Be$ interactions
as a function of \xf\  compared with: (a) the E771 ($p Si$) [10] and E789 
($p Au$) [9] results
(each data sample is normalized by an integral of
the distribution over the $0.0<x_F<0.135$ range); (b) the E789
($p Be$) and ($p Cu$) [8] results
(each data sample is normalized by an integral of
the distribution over the $0.3<x_F<0.55$ range).

\item Figure~10.  Ratios of differential distributions as functions of \xf\
for \jpsi\
production in: 
 (a) $pp$  and $p Be$ interactions, and (b) $p Cu$ and $p Be$ interactions.
  The open and full circles represent the 530~GeV/$c$ and 800~GeV/$c$ data,
  respectively.
  The integrals of the input distributions are normalized to unity.

\item Figure~11.  Center-of-mass energy dependence of the average \jpsi\ 
transverse momentum for nucleon-nucleus interactions.

\item Figure~12.  Ratios of differential distributions as functions of
\pts\ for $J/\psi$
production in: 
 (a) $pp$  and $p Be$ interactions, and (b) $p Cu$ and $p Be$ interactions.
  The open and full circles represent the 530~GeV/$c$ and 800~GeV/$c$ data,
  respectively.
  The integrals of the input distributions are normalized to unity.

\end{list}

\newpage

\begin{figure}
\vspace{-0.425in}
\vbox
{
\centerline{\epsfxsize=6.0in
\epsfbox{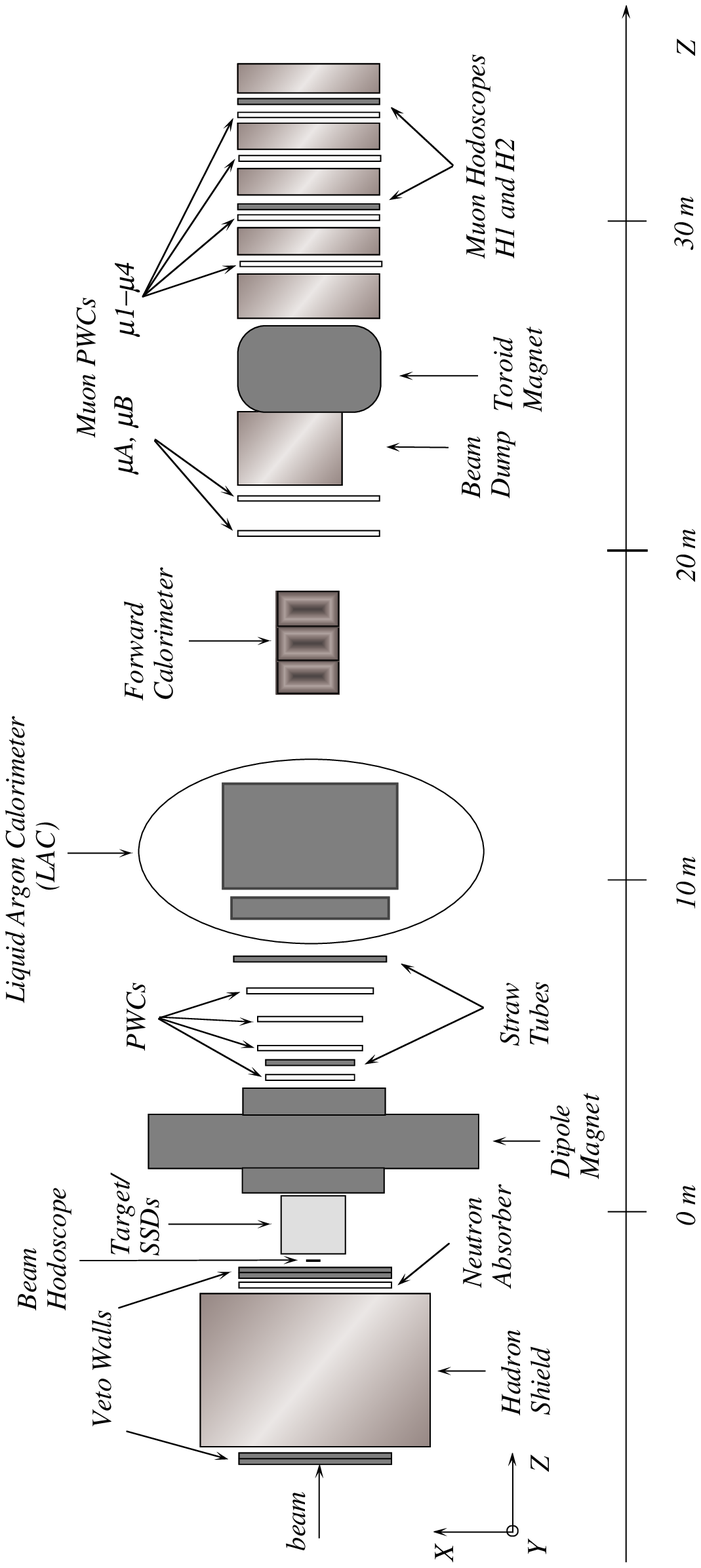}}
\vspace{0.425in}
\caption{
 Plan view of the Fermilab Meson West Spectrometer as
configured for the
1991 fixed target run.}
}
\label{Fig1}
\end{figure}

\newpage
\begin{figure}
\vspace{-0.125in}
\vbox
{
\centerline{\epsfxsize=6.0in
\epsfbox{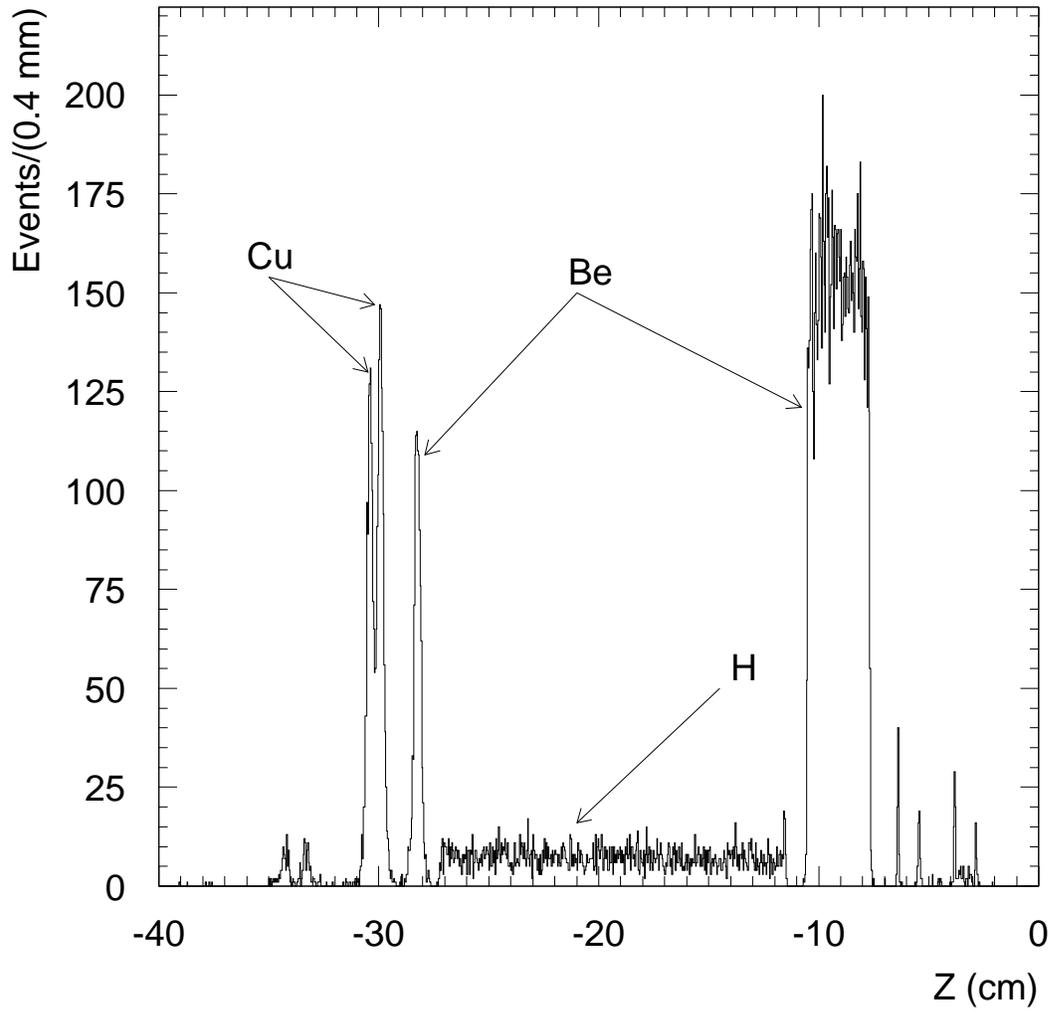}}
\vspace{0.825in}
\caption{
Primary vertex Z-coordinate distribution for events containing
dimuons with reconstructed masses in the \jpsi\ mass range. Unmarked spikes in
the distribution upstream of the $Cu$ target and downstream of the $Be$ target
are due to interactions in the silicon-strip detectors.}
}
\label{Fig2}
\end{figure}

\newpage
\begin{figure}
\vspace{-0.125in}
\vbox
{
\centerline{\epsfxsize=5.0in
\epsfbox{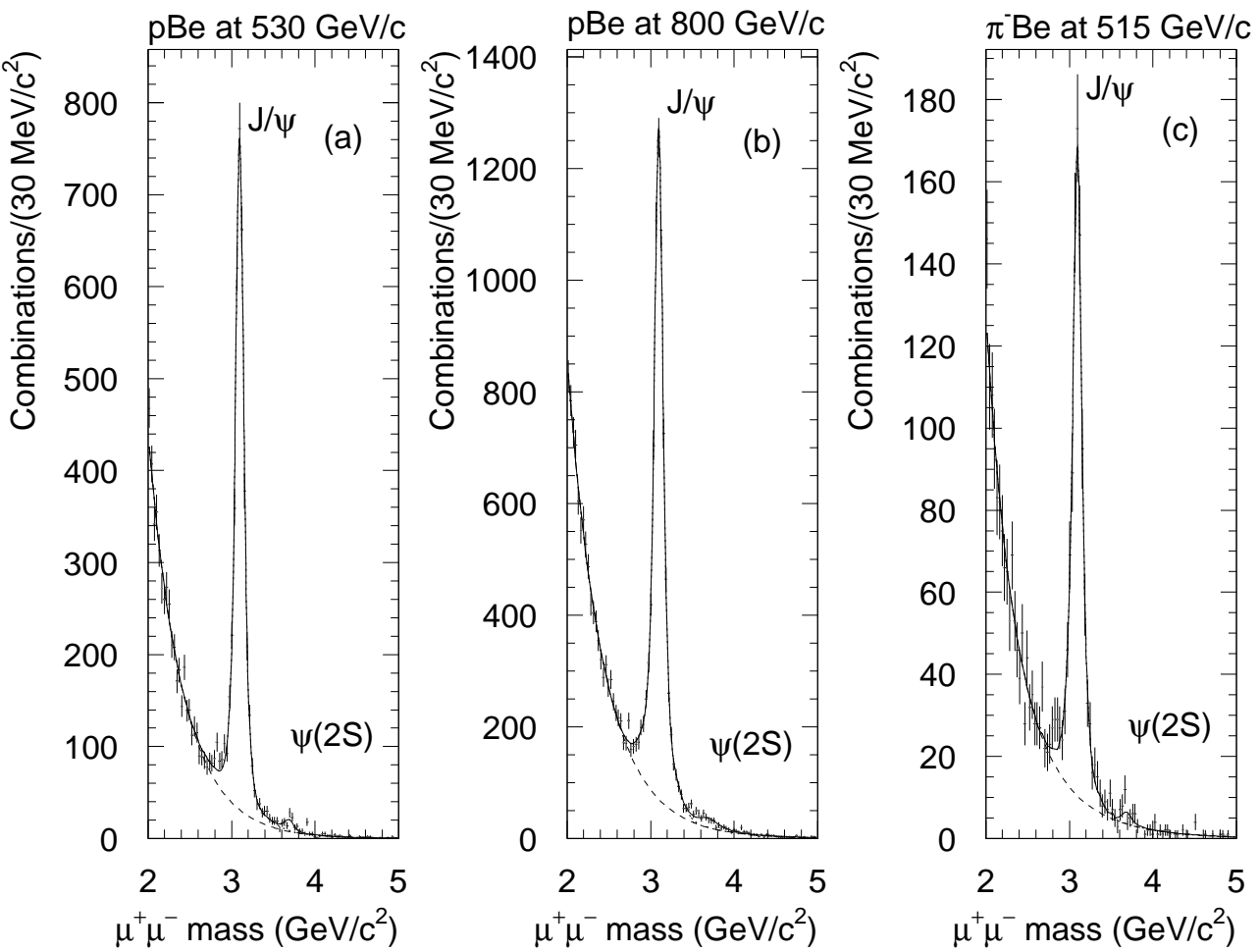}}
\vspace{0.5in}
\vspace{0.825in}
\caption{
 The invariant mass distributions of $\mu^+\mu^-$ pairs
for: (a) 530~GeV/$c$ incident protons, (b) 800~GeV/$c$ incident protons,
and (c) 515~GeV/$c$ incident $\pi^-$. The solid curve in each plot is a fit to
the data; the dashed curve shows the
background contribution.}
}
\label{Fig3}
\end{figure}

\newpage
\begin{figure}
\vspace{-0.125in}
\vbox
{
\centerline{\epsfxsize=5.0in
\epsfbox{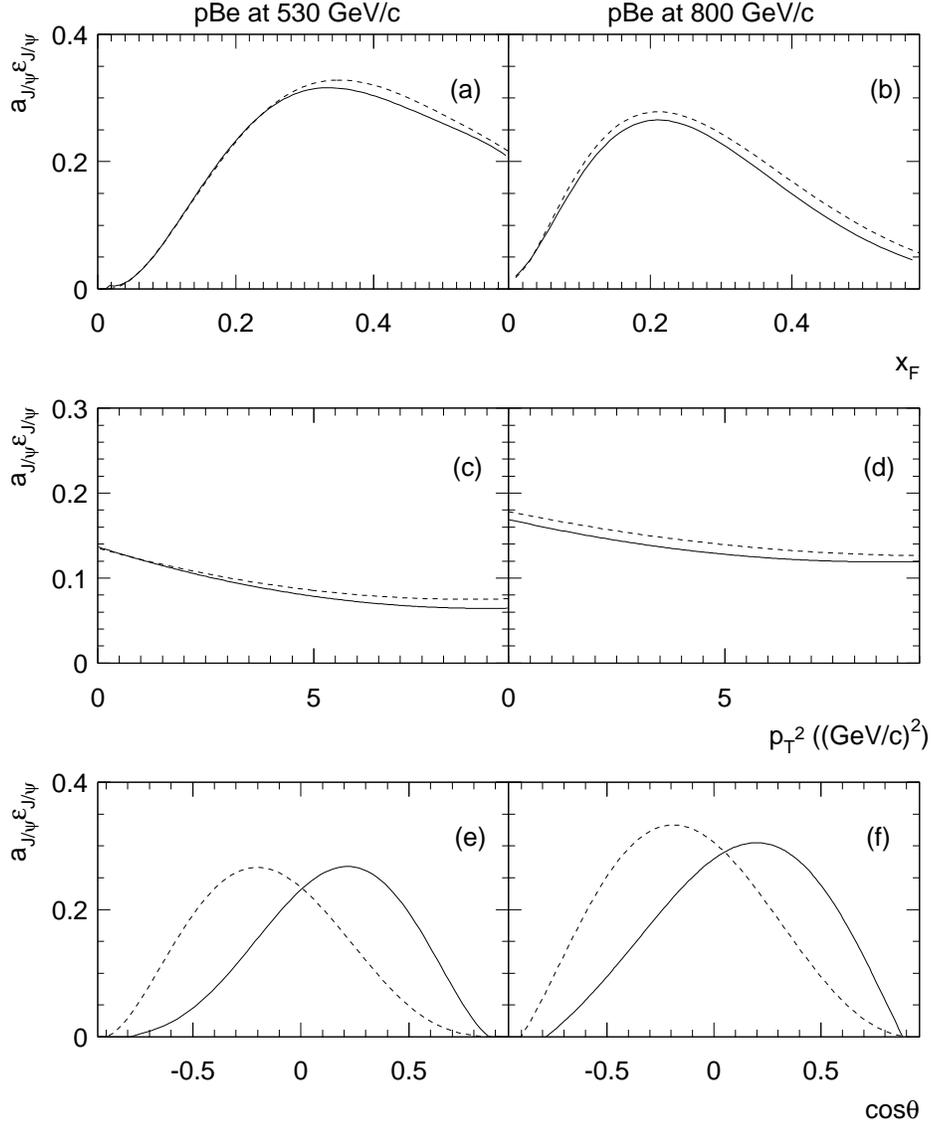}}
\vspace{0.825in}
\caption{
Product of geometrical acceptance and  reconstruction
efficiency for detection of \psimumu\
as a function of: (a,b) \xf , (c,d) \pts , and (e,f) $cos \theta$
for 530~GeV/$c$
(left column) and 800~GeV/$c$ (right column) incident protons.
The solid and dashed curves represent the results for the positive and
negative toroid
polarity, respectively.}
}
\label{Fig4}
\end{figure}

\newpage
\begin{figure}
\vspace{-0.425in}
\vbox
{
\centerline{\epsfxsize=4.5in
\epsfbox{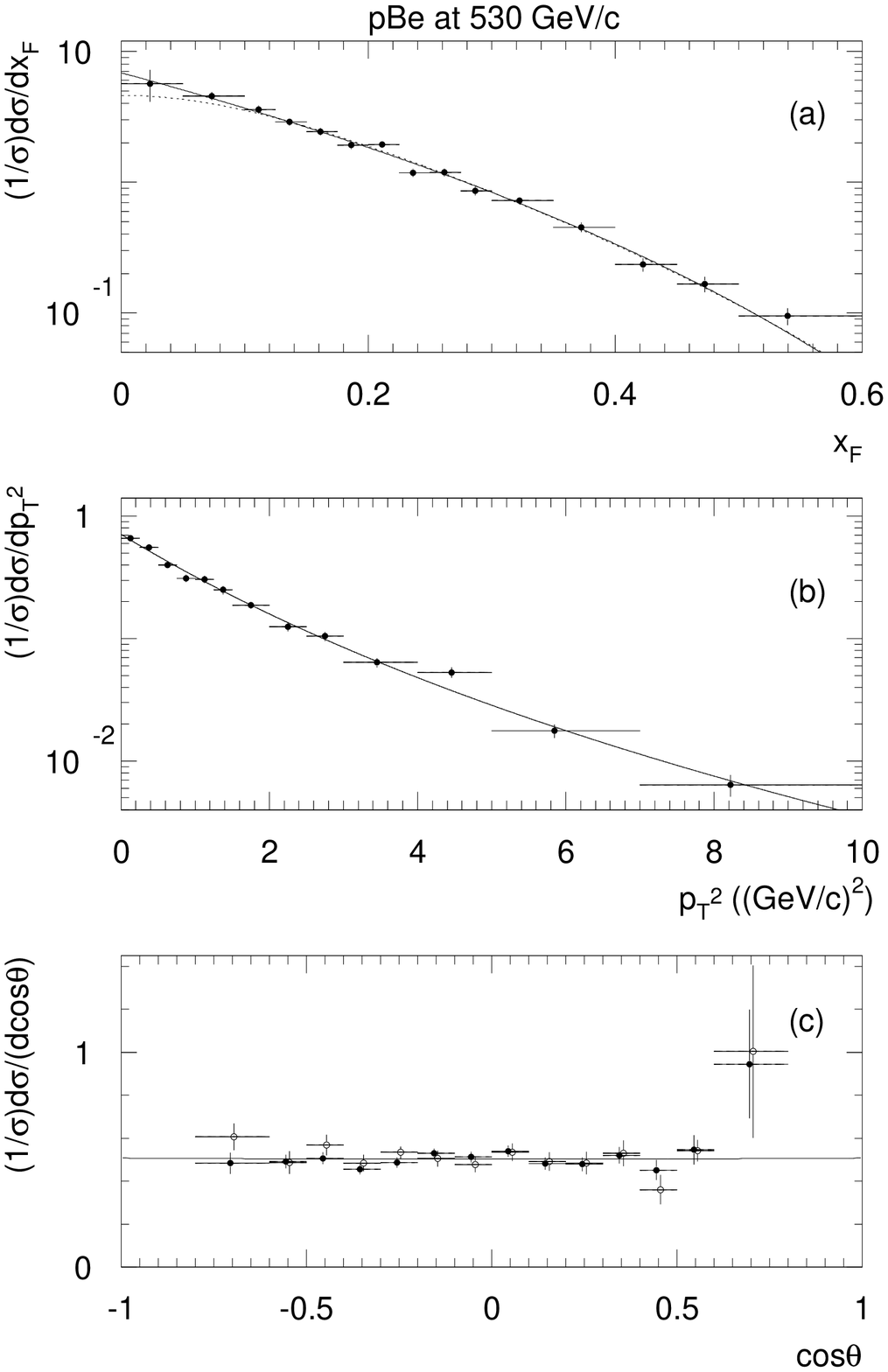}}
\vspace{0.425in}
\caption{
Differential distributions for \jpsi\ production
as functions of: (a) \xf  , (b)
\pts\  ((GeV/$c)^{2}$), and  (c) $cos \theta$  for $p Be$ interactions at 
530~GeV/$c$. The $cos \theta$ distributions are shown separately for each
toroid magnet
polarity.
The integrals of the distributions in (a) and (b) are normalized to unity.
The integrals of the $cos \theta$ distributions within the
-0.6$<cos \theta <0.6 $ range are normalized to 0.6.
The solid and dashed curves in (a) represent empirical fits to the data using
Eqs.~(4.1) and (4.2), respectively.
The solid curves in (b) and (c) represent fits using the
functions shown in Eqs.~(4.4) and (4.5), respectively.
}
}
\label{Fig5}
\end{figure}

\newpage
\begin{figure}
\vspace{-0.325in}
\vbox
{
\centerline{\epsfxsize=4.5in
\epsfbox{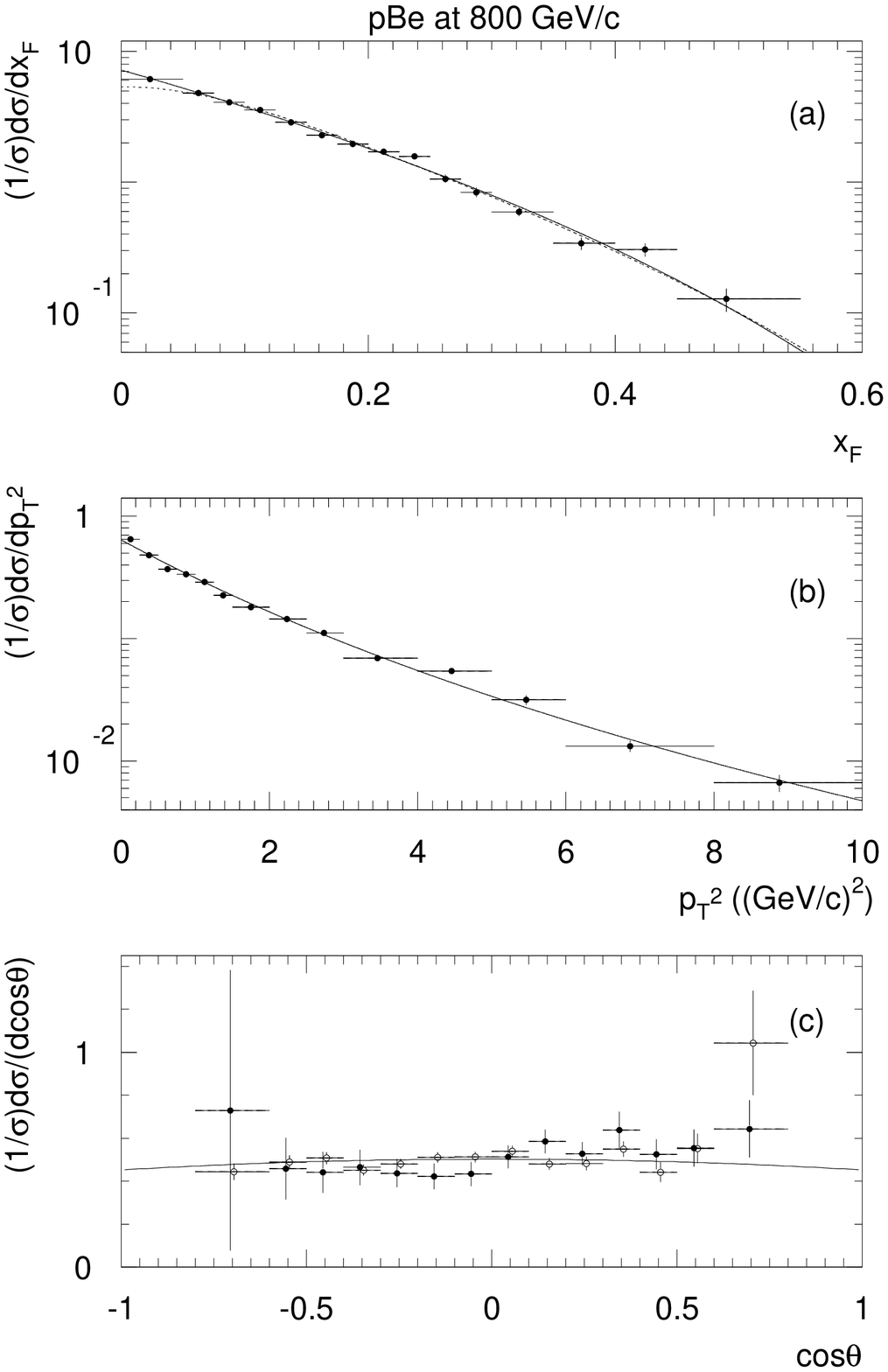}}
\vspace{0.425in}
\caption{
Differential distributions for \jpsi\ production as functions of: (a) \xf  ,
(b) \pts\  ((GeV/$c)^{2}$), and  (c) $cos \theta$  for $p Be$ interactions at 
800~GeV/$c$. The $cos \theta$ distributions are shown separately for each
toroid magnet
polarity.
The integrals of the distributions in (a) and (b) are normalized to unity.
The integrals of the $cos \theta$ distributions within the 
-0.6$<cos \theta <$0.6 range
are normalized to 0.6.
The solid and dashed curves in (a) represent empirical fits to the data using
Eqs.~(4.1) and (4.2), respectively.
The solid curves in (b) and (c) represent fits using the
functions shown in Eqs.~(4.4) and (4.5), respectively.
}
}
\label{Fig6}
\end{figure}

\newpage
\begin{figure}
\vspace{-0.325in}
\vbox
{
\centerline{\epsfxsize=4.5in
\epsfbox{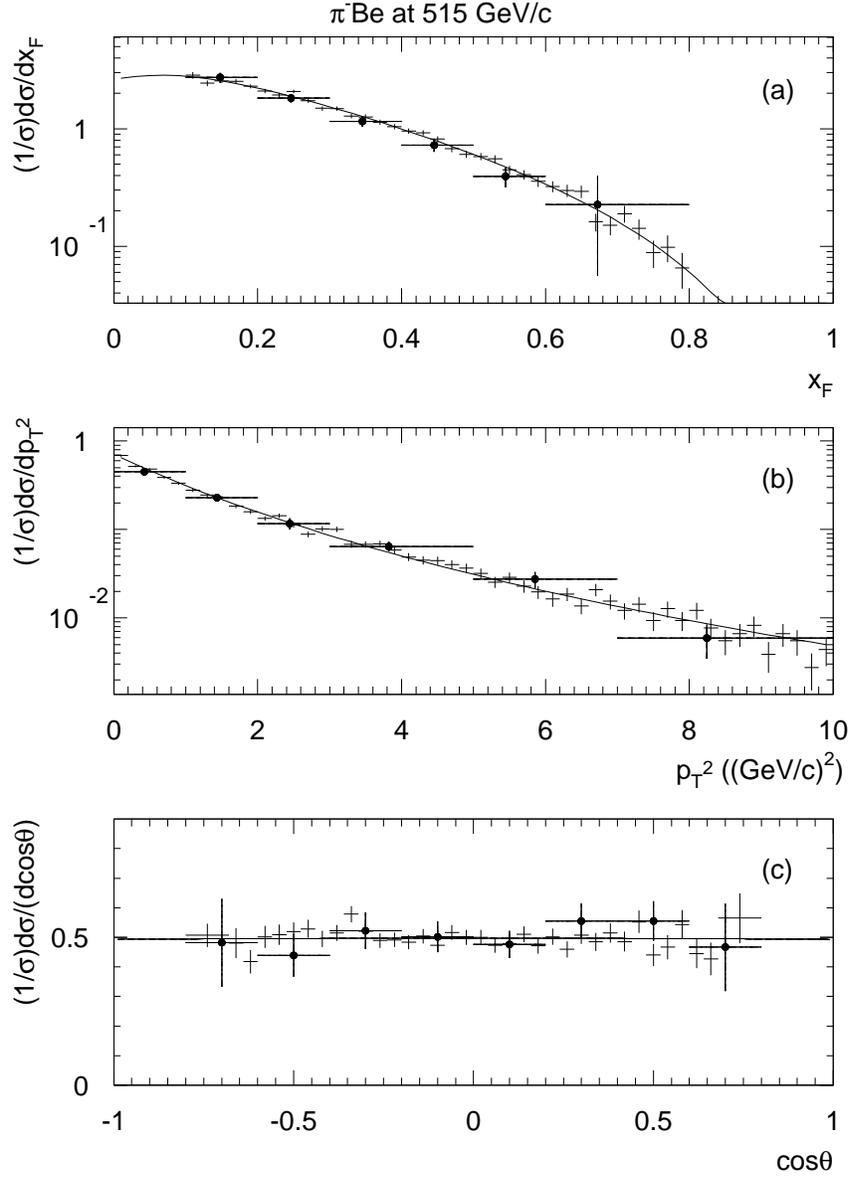}}
\vspace{0.425in}
\caption{ 
Differential distributions for $J/\psi$ production
as functions of: (a) \xf  , (b)
\pts\  ((GeV/$c)^{2}$), and  (c) $cos \theta$  for $\pi ^- Be$ interactions at
515~GeV/$c$.
The integrals of the distributions in (a) and (b) are normalized to unity.
The integral of the $cos \theta$ distribution within the
-0.6$<cos \theta <$0.6 range
is normalized to 0.6.
The data of Ref.[6] ~are shown for comparison. 
}
}
\label{Fig7}
\end{figure}

\newpage
\begin{figure}
\vspace{-0.125in}
\vbox
{
\centerline{\epsfxsize=6.0in
\epsfbox{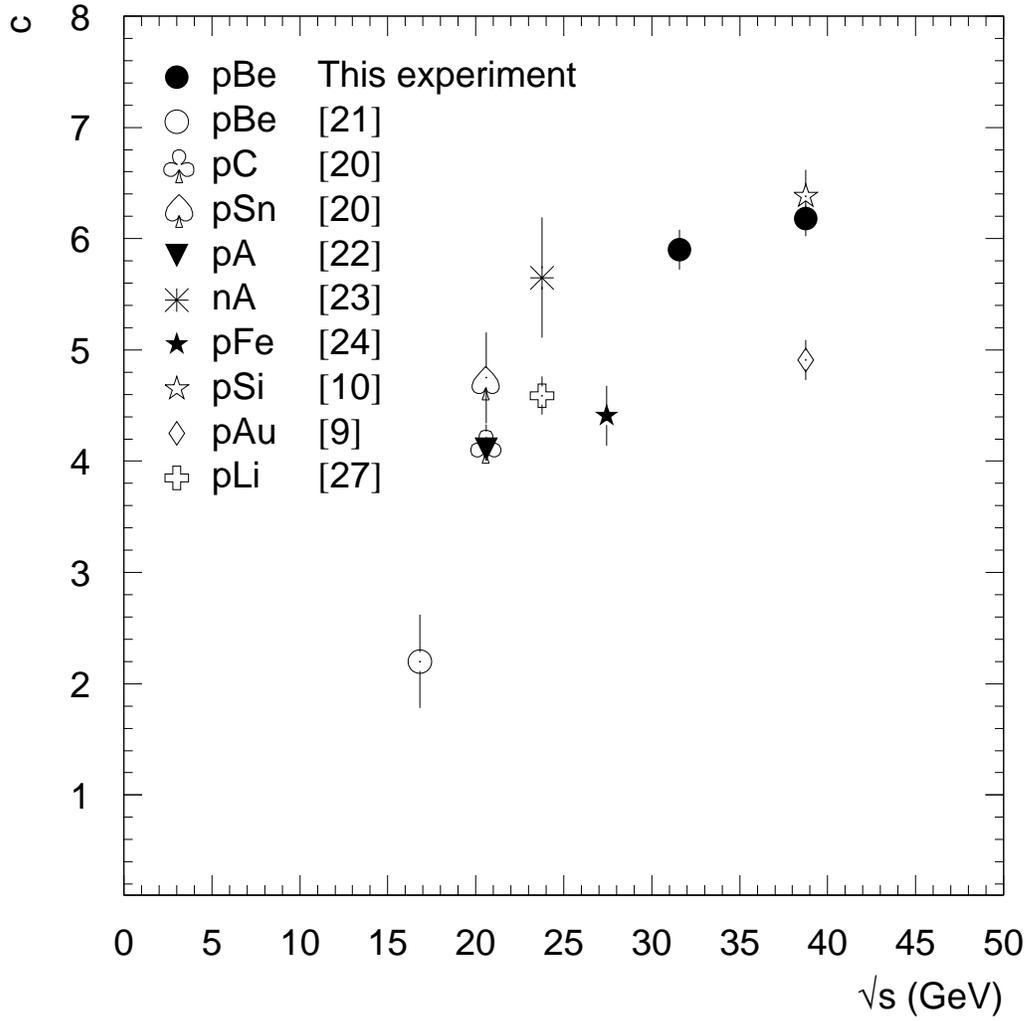}}
\vspace{0.825in}
\caption{ 
Center-of-mass energy dependence of the exponent $c$ in  the
parametrization  of the  \xf\ distributions for \jpsi\ production in
nucleon-nucleus
interactions. The \xf\  distributions for Refs.~[20,21,22] were converted into
the d$\sigma$/d\xf\ form from the quoted Lorentz invariant cross sections
$E d\sigma$/d\xf\ using the average value of the \jpsi\ transverse
momentum at a given center-of-mass energy.
}
}
\label{Fig8}
\end{figure}

\newpage
\begin{figure}
\vspace{-0.525in}
\vbox
{
\centerline{\epsfxsize=5.5in
\epsfbox{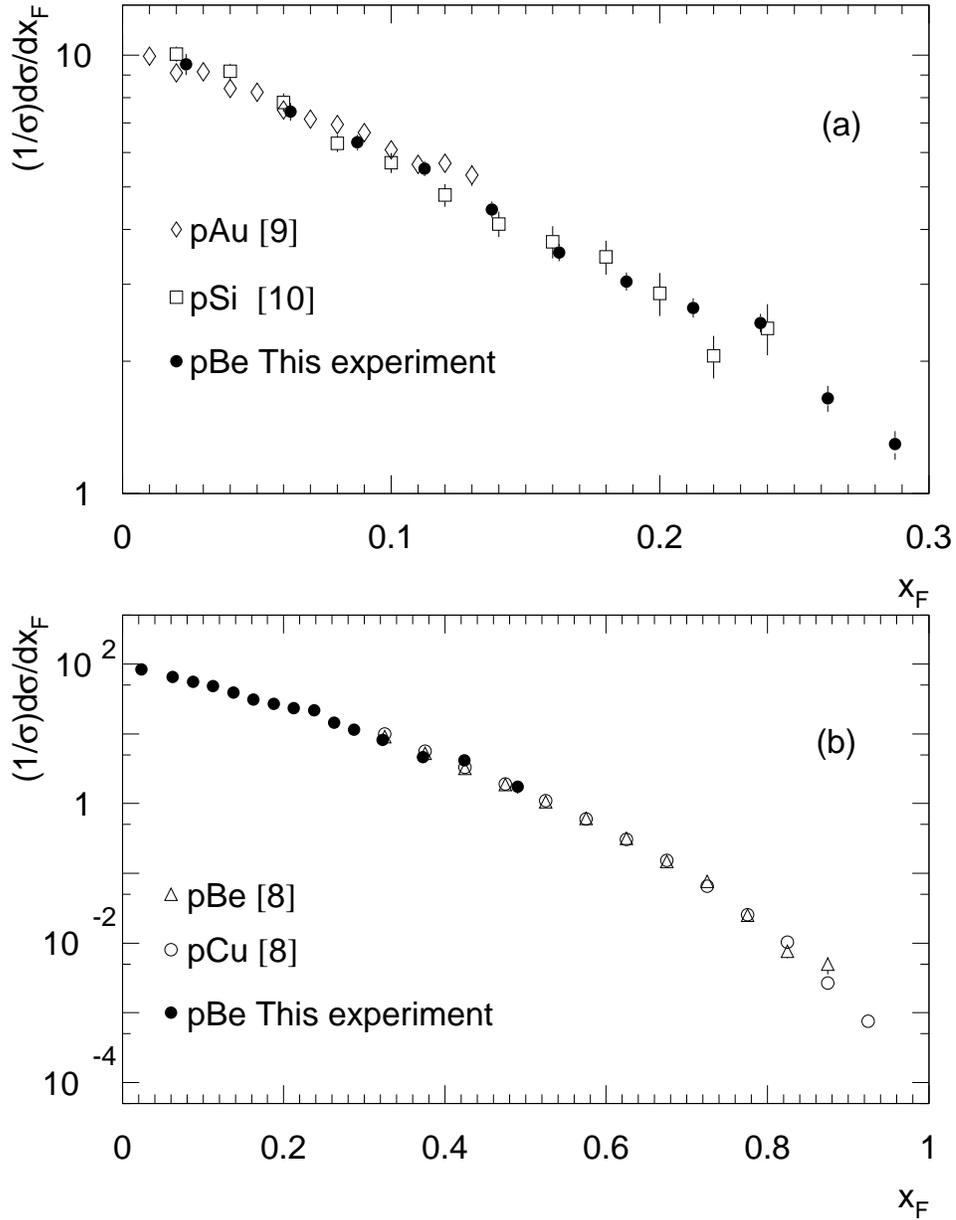}}
\vspace{0.425in}
\caption{ 
 Differential distribution for $J/\psi$ production in 
800~GeV/$c$ $p Be$ interactions
as a function of \xf\  compared with: (a) the E771 ($p Si$) [10] and E789
($p Au$) [9] results
(each data sample is normalized by an integral of
the distribution over the $0.0<x_F<0.135$ range); (b) the E789
($p Be$) and ($p Cu$) [8] results
(each data sample is normalized by an integral of
the distribution over the $0.3<x_F<0.55$ range).
}
}
\label{Fig9}
\end{figure}

\newpage
\begin{figure}
\vspace{-0.425in}
\vbox
{
\centerline{\epsfxsize=6.0in
\epsfbox{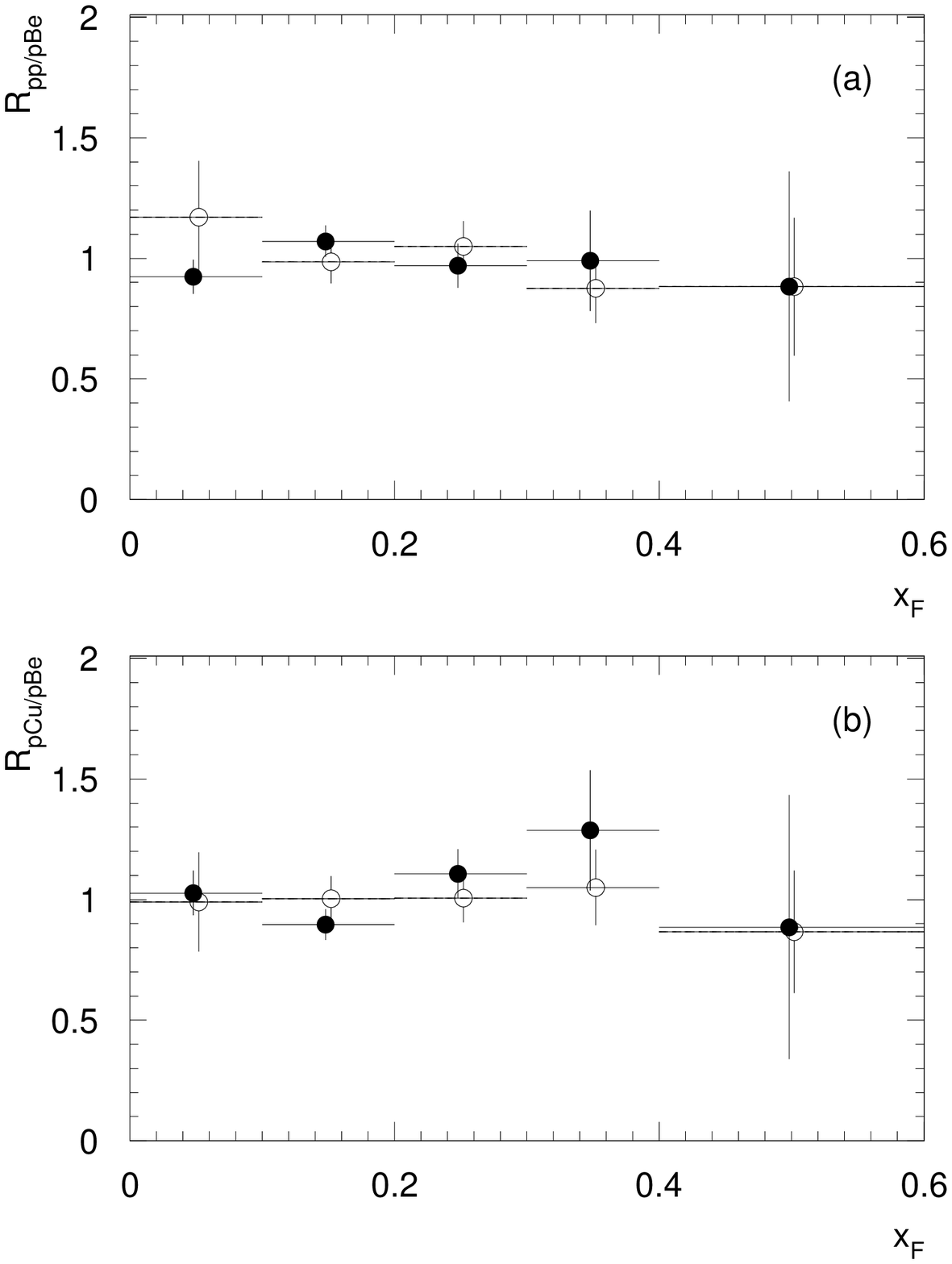}}
\vspace{0.425in}
\caption{
 Ratios of differential distributions as functions of \xf\ for \jpsi\
production in:
 (a) $pp$  and $p Be$ interactions, and (b) $p Cu$ and $p Be$ interactions.
  The open and full circles represent the 530~GeV/$c$ and 800~GeV/$c$ data,
  respectively.
  The integrals of the input distributions are normalized to unity.
}
}
\label{Fig10}
\end{figure}

\newpage
\begin{figure}
\vspace{-0.325in}
\vbox
{
\centerline{\epsfxsize=6.0in
\epsfbox{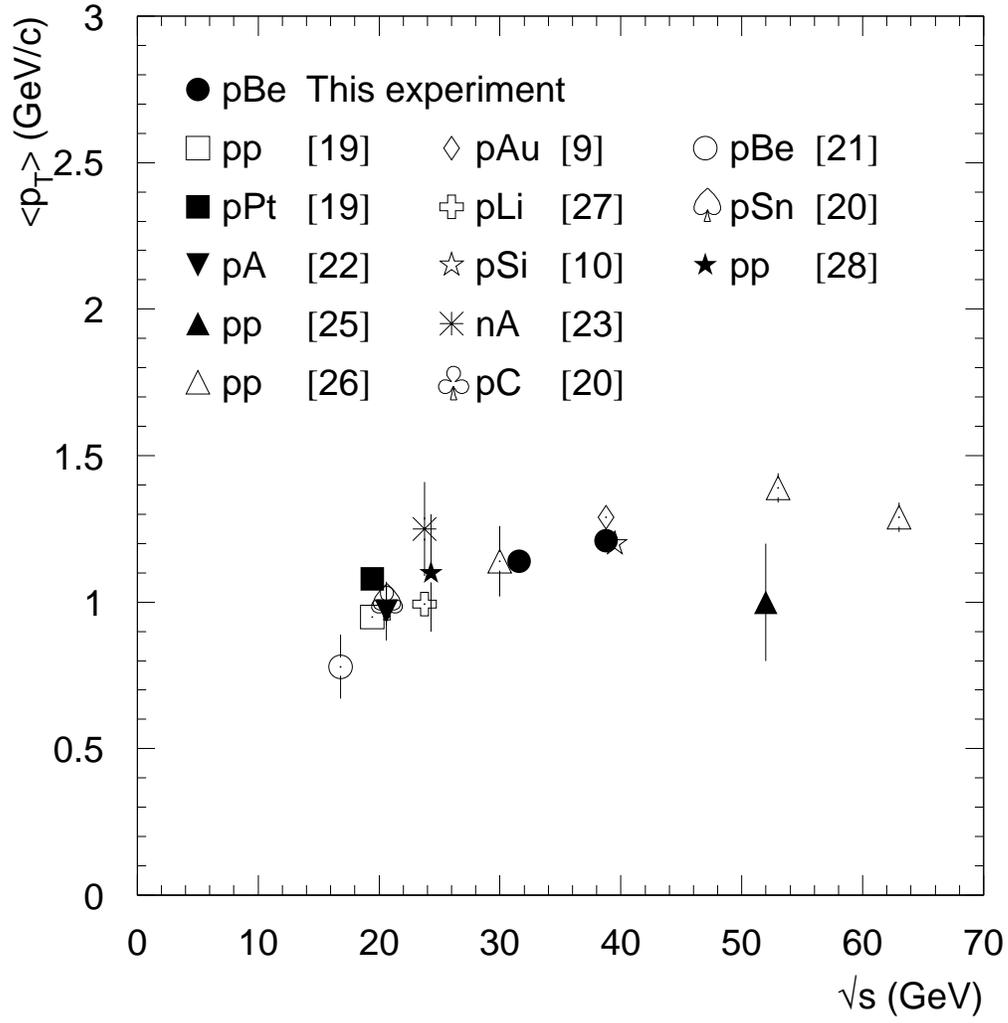}}
\vspace{0.425in}
\caption{
 Center-of-mass energy dependence of the average \jpsi\
transverse momentum for nucleon-nucleus interactions.
}
}
\label{Fig11}
\end{figure}

\newpage
\begin{figure}
\vspace{-0.425in}
\vbox
{
\centerline{\epsfxsize=6.0in
\epsfbox{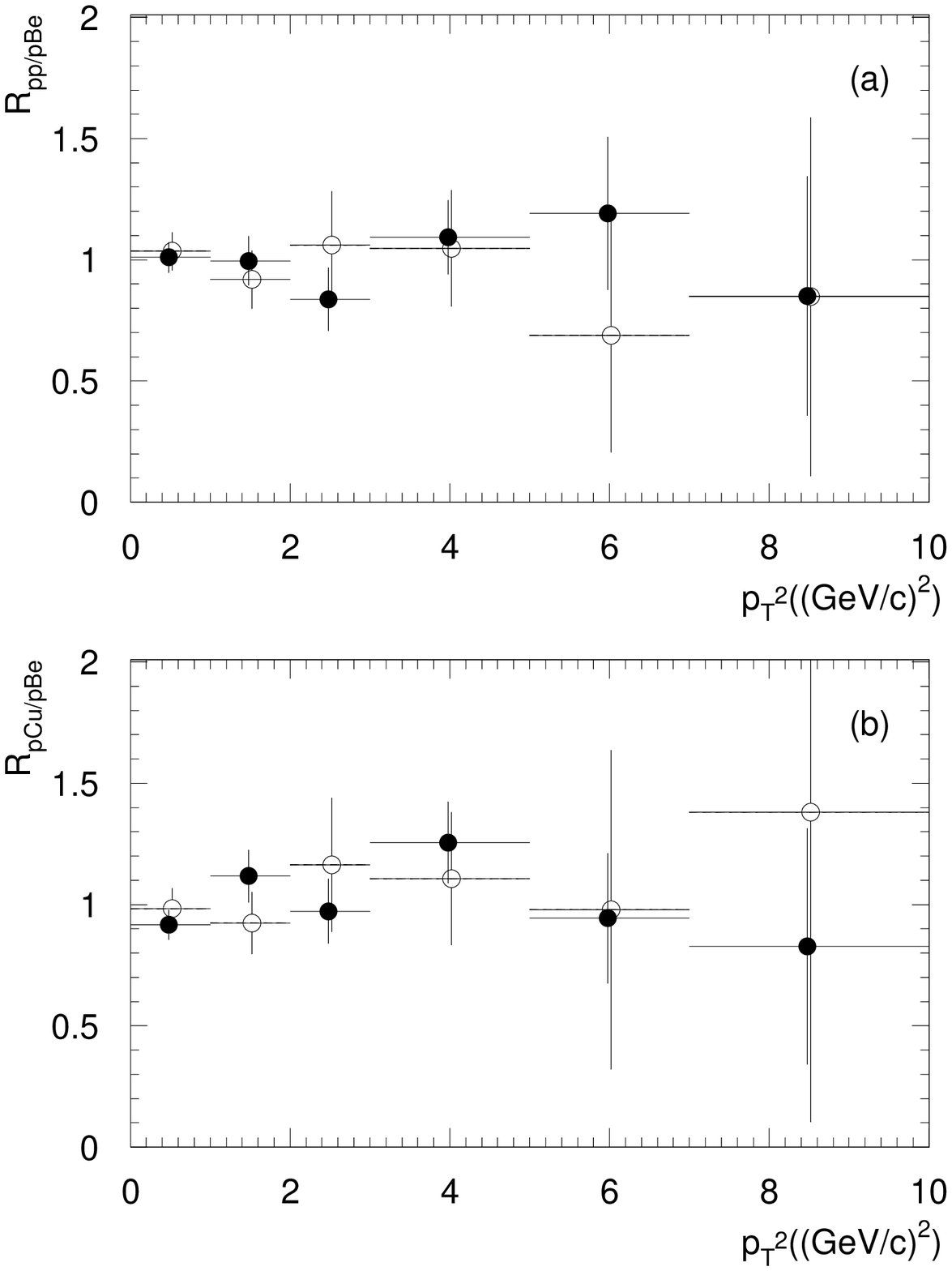}}
\vspace{0.425in}
\caption{
 Ratios of differential distributions as functions of \pts\ for $J/\psi$
production in:
 (a) $pp$  and $p Be$ interactions, and (b) $p Cu$ and $p Be$ interactions.
  The open and full circles represent the 530~GeV/$c$ and 800~GeV/$c$ data,
  respectively.
  The integrals of the input distributions are normalized to unity.
}
}
\label{Fig12}
\end{figure}

\end{document}

%% file: list_of_authors.tex
%
\author{                                                                        
A.~Gribushin,$^{4}$
V.~Abramov,$^{2}$
Yu.~Antipov,$^{2}$
B.~Baldin,$^{2}$
R.~Crittenden,$^{4}$
C.~Davis,$^{5}$
L.~Dauwe,$^{6}$
S.~Denisov,$^{2}$
A.~Dyshkant,$^{2}$
A.~Dzierba,$^{4}$
V.~Glebov,$^{2}$
H.~Goldberg,$^{3}$
R.~Jesik,$^{4}$
V.~Koreshev,$^{2}$
J.~Krider,$^{1}$
A.~Krinitsyn,$^{2}$
R.~Li,$^{4}$
S.~Margulies,$^{3,\ddag}$
T.~Marshall,$^{4}$ 
J.~Martin,$^{4}$
H.~Mendez,$^{3}$
A.~Petrukhin,$^{2}$
J.~Solomon,$^{3}$
V.~Sirotenko,$^{2}$
P.~Smith,$^{4}$
T.~Sulanke,$^{4}$
R.~Sulyaev,$^{2,\ddag}$
F.~Vaca,$^{3}$
A.~Ziemi\' nski$^{4}$
\\
\vskip 0.25cm                                                                   
\centerline{(E672 Collaboration)}                                               
\vskip 0.25cm                                                                   
L.~Apanasevich,$^{9}$
J.~Bacigalupi,$^{7}$
M.~Begel,$^{13}$
S.~Blusk,$^{12}$
C.~Bromberg,$^{9}$
P.~Chang,$^{10}$
B.~Choudhary,$^{8}$
W.~H.~Chung,$^{12}$
L.~de~Barbaro,$^{13}$
W.~D\l ugosz,$^{10}$
J.~Dunlea,$^{13}$
E.~Engels,~Jr.,$^{12}$
G.~Fanourakis,$^{13}$
G.~Ginther,$^{13}$
K.~Hartman,$^{11}$
J.~Huston,$^{9}$
V.~Kapoor,$^{8}$
C.~Lirakis,$^{10}$
S.~Mani,$^{7}$
J.~Mansour,$^{13}$
A.~Maul,$^{9}$
R.~Miller,$^{9}$
G.~Osborne,$^{13}$
D.~Pellett,$^{7}$
E.~Pothier,$^{10}$
R.~Roser,$^{13}$
P.~Shepard,$^{12}$
D.~Skow,$^{1}$
P.~Slattery,$^{13}$
L.~Sorrell,$^{9}$
D.~Striley,$^{10}$
N.~Varelas,$^{13}$
D.~Weerasundara,$^{12}$
C.~Yosef,$^{9}$
M.~Zieli\' nski,$^{13}$
V.~Zutshi$^{8}$
\\
\vskip 0.25cm                                                                   
\centerline{(E706 Collaboration)}                                               
\vskip 0.25cm                                                                   
}                                                                               
\address{                                                                       
\centerline{$^{1}$Fermi National Accelerator Laboratory, Batavia,              
                   Illinois 60510}                                              
\centerline{$^{2}$Institute for High Energy Physics, Serpukhov, Russia} 
\centerline{$^{3}$University of Illinois at Chicago, Chicago, Illinois 60607}             
\centerline{$^{4}$Indiana University, Bloomington, Indiana 47405}              
\centerline{$^{5}$University of Louisville, Louisville, Kentucky 40292}
\centerline{$^{6}$University of Michigan at Flint, Flint, Michigan 48502}
\centerline{$^{7}$University of California-Davis, Davis, California 95616}
\centerline{$^{8}$University of Delhi, Delhi, India 110007}                       
\centerline{$^{9}$Michigan State University, East Lansing, Michigan 48824}     
\centerline{$^{10}$Northeastern University, Boston, Massachusetts  02115}
\centerline{$^{11}$Pennsylvania State University, University Park, 
		   Pennsylvania 16802}
\centerline{$^{12}$University of Pittsburgh, Pittsburgh, Pennsylvania 15260}
\centerline{$^{13}$University of Rochester, Rochester, New York 14627}          
\centerline{$^{\ddag}$Deceased}
}                                                                               

%% file: e672_symbols.tex
\def\mupmum{$\mu^+ \mu^-$}
\def\qq{$q\bar{q}$}
\def\pim{$\pi ^{-}$}
\def\Km{$K^{-}$}
\def\K0{$K^\circ$}
\def\Lb{$\Lambda _{b}$}
\def\jpsi{$J/ \psi$} 
\def\psip{$\psi(2S)$}
\def\psimumu{$J/\psi \rightarrow \mu^+ \mu^- $}
\def\psipipi{$J/ \psi \pi^+ \pi^-$}
\def\psipi{$J/\psi \pi^{\pm}$}
\def\pt{$p_T$}
\def\pts{$p_T^2$}
\def\dsigdpt{$d\sigma/dp_T$}
\def\avpt{${\langle {p_T} \rangle}$}
\def\sqs{$\sqrt{s}$}
\def\xf{$x_F$}
\def\costh{$cos\theta$}
\def\cossqth{cos$^{2}\theta$}

\def\B{$B^\circ$}
\def\pp{$p\bar{p}$}
\def\jpsig{${J/ \psi - \gamma}$}
\def\jpsiee{${J/ \psi e^+e^-}$}
\def\deltam{$\Delta M$}
\def\fc{$f_\chi$}
\def\fb{$f_b$}
\def\deg#1{\mbox{${#1}^{\circ}$}}
\def\bbar{$b\bar{b}$}
\def\cc{$c\bar{c}$}
\def\r12{$R(\chi_{c1} / \chi_{c2})$}
\def\rcj{$R(\chi / \psi)$}
\def\rprime{$R(\psi / \psi\prime)$}
\def\xhic{$\chi_{c}$} 
\def\xhio{$\chi_{c1}$} 
\def\xhit{$\chi_{c2}$} 
\def\gg-chic-jpsigamma{$gg \rightarrow \chi_{c} X \rightarrow J/ \psi + \gamma$}
\def\ctop{$\chi_c  \rightarrow J/ \psi + \gamma, J/\psi \rightarrow \mu \mu$}
\def\BB-jpsi+x{$B \bar{B} \rightarrow J/ \psi X $}
\def\mumug{$\mu \mu \gamma $}
\def\mdif{$M(\mu \mu \gamma) - M(\mu^+ \mu^-)$}
\def\D0{D\O\ }
\def\Rphi{$R-\phi$}
\def\als{$\alpha_s$}
\def\emu{$e\mu$}
\def\mm{$\mu m$}
\def\BDmuX{$B \rightarrow D \mu X $}
\def\BDsmu{$B \rightarrow D^{\star} \mu X $}
\def\thirdo{$O (\alpha _s^3)$} 
\def\seco{$O (\alpha _s^2)$} 
\def\ptpsi{$p_T^{\psi}$}
\def\ptmin{$p_T^{min}$}
\def\ptrel{$p_T^{rel}$}
\def\poverm{${p_T}^2/{m_c}^2$}
\def\ptmu{$p_T^{\mu}$}
\def\etamu{${\eta} ^{\mu}$}
\def\etae{$\abs{\eta} < 0.6$}
\def\etaf{$\abs{\eta} < 1.0$}
\def\dphi{$\Delta\phi$}
\def\deta{$\Delta\eta$}
\def\dthe{$\Delta\theta$}
\def\dr{$\Delta$$R$}
\def\pppsi{$p \bar{p} \rightarrow J/ \psi X \rightarrow \mu^+ \mu^- X$}
\def\mass{$M_{\mu\mu}$}
\def\ipb{$pb^{-1}$}
\def\inb{$nb^{-1}$}
\def\jetet{$E_{T}^{jet}$}
\def\ptmumu{$p_{T}^{\mu\mu}$}
\def\Dzer{$D^\circ$}
\def\Ds{$D^\star$}
\def\Dbar{$\bar{D}$}
\def\fb{$f_{b}$}
\def\fc{$f_{\chi}$}
\def\Demu{$\Delta_{e\mu}$}
\def\Dphiemu{$\Delta\phi_{e\mu}$}
\def\Memu{$M_{e\mu}$}
\def\dr{$\Delta R$}
\def\mumug{$M_{\mu\mu\gamma}$}
\def\mumu{$M_{\mu\mu}$}
\def\Ldtapp{$\int Ldt \approx$}
\def\Ete{$E_T^{e}$}
\def\Etjet{$E_T^{jet}$}



%% file: e672_psi97_prd_gg.bbl
\begin{references}

\bibitem{CDFD0} F.~Abe {\sl et al.},
{Phys.~Rev.~Lett.} {\bf 79}, 572 (1997); 
S.~Abachi {\sl et al.},
{Phys.~Lett.} {\bf B370}, 239 (1996).

\bibitem{Braat} E.~Braaten and S.~Fleming,
{Phys.~Rev.~Lett.}
{\bf 74}, 3327 (1995);
M.~Cacciari {\sl et al.},
{Phys.~Lett.~} {\bf B356}, 553 (1995); 
W.~Tang and M.~Vanttinen, {Phys.~Rev.~D} {\bf 53}, 4851 (1996);
P. Cho and A.~K.~Leibovich,
{Phys.~Rev.~D} {\bf 53}, 150 and 6203 (1996).

\bibitem{Gupta} 
S.~Gupta and K.~Sridhar,
{Phys.~Rev.~D}
{\bf 54}, 5545 (1996); {\bf 55}, 2650 (1997);
M.~Beneke and I.~Z.~Rothstein, {Phys.~Rev.~D} {\bf 54}, 2005 (1996); {\bf 54},
7082 (E) (1996).

\bibitem{Baier} R.~Baier and R.~Ruckl,
{Z.~Phys.~C} {\bf 19}, 251 (1983);
 M.~Vanttinen {\sl et al.},
{Phys.~Rev.~D}
{\bf 51}, 3332 (1995).

\bibitem{Halzen} J.~F.~Amundson {\sl et al.},
{Phys. Lett.} {\bf B390}, 323 (1997);

\bibitem{Grib} A.~Gribushin {\sl et al.},
{Phys.~Rev.~D}
{\bf 53}, 4723 (1996).


\bibitem{Kore} V.~Koreshev {\sl et al.},
{Phys.~Rev.~Lett.} {\bf 77}, 4294 (1996).

\bibitem{Kowitt} M.~S.~Kowitt {\sl et al.},
{Phys.~Rev.~Lett.}
{\bf 72}, 1318 (1994).

\bibitem{Esen} M.~H.~Schub {\sl et al.},
{Phys.~Rev.~D}
{\bf 52}, 1307 (1995); {Phys.~Rev.~D}
{\bf 53}, 570 (E) (1996).

\bibitem{Esso} T.~Alexopoulos {\sl et al.},
{Phys.~Rev.~D}
{\bf 55}, 3927 (1997).

\bibitem{Blusk} L.~Apanasevich {\sl et al.},
{Phys.~Rev.~D}
{\bf 56}, 1391 (1997).

\bibitem{Striley} D.~Striley,
Ph.D. Thesis, University of Missouri-Columbia, 1996 (unpublished).

\bibitem{PDG} Particle Data Group, 
{Eur.~ Phys.~ J.} 
{\bf C 3}, 1 (1998).

\bibitem{Abramov} V.~Abramov {\sl et al.},
``Properties of $J/\psi$ Production in $\pi^- Be$
and $p Be$ Collisions at 530~GeV/$c$'',
FERMILAB-Pub-91/62-E 1991 (unpublished). The cross sections quoted in this
preprint must be increased by a factor 1.2 to account for a correction omitted
in the luminosity calculations.

\bibitem{AZWaw} D.~M.~Alde {\sl et al.},
{Phys.~Rev.~Lett.}
{\bf 66}, 133 (1991); M.~J.~Leitch {\sl et al.},
{Phys.~Rev.~D}
{\bf 52}, 4251 (1995); M.~J.~Leitch, Proceedings of the Quark Matter '99
Conf., May 1999 - in press.

\bibitem{Likho} V.~G.~Kartvelishvili and A.~K.~Likhoded,
{Sov. J. Nucl. Phys.}
{\bf 39}, 298 (1984).

\bibitem{Kaplan} D.~M.~Kaplan {\sl et al.}, 
{Phys.~Rev.~Lett.} 
{\bf 40}, 435 (1978).

\bibitem{Akerlof} C.~Akerlof {\sl et al.}, 
{Phys.~Rev.~D} 
{\bf 48}, 5057 (1993) and references therein.

\bibitem{Badier} J.~Badier {\sl et al.},
{Z.~Phys.~C} 
{\bf 20}, 101 (1983).

\bibitem{Brans} J.~G.~Branson  {\sl et al.},
{Phys.~Rev.~Lett.} 
{\bf 38}, 1331 (1977).

\bibitem{Anders} K.~J.~Anderson {\sl et al.},
{Phys.~Rev.~Lett.}
{\bf 37}, 799 (1976).

\bibitem{Andtwo} K.~J.~Anderson {\sl et al.},
{Phys.~Rev.~Lett.} 
{\bf 42}, 944 (1979).

\bibitem{Binkley} M.~Binkley {\sl et al.},
{Phys.~Rev.~Lett.} 
{\bf 37}, 574 (1976).

\bibitem{Siskind} E.~J.~Siskind {\sl et al.},
{Phys.~Rev.~D} 
{\bf 21}, 628 (1980).

\bibitem{Nagy} E.~Nagy {\sl et al.},
{Phys.~Lett.}
{\bf 60B}, 96 (1975).

\bibitem{Kourk} C.~Kourkomelis {\sl et al.},
{Phys.~Lett.}
{\bf 91B}, 481 (1980).

\bibitem{Anton} C.~Antoniazzi  {\sl et al.},
{Phys.~Rev.~D}
{\bf 46}, 4828 (1992).

\bibitem{Morel} C.~Morel  {\sl et al.},
{Phys.~Lett. B}
{\bf 252}, 505 (1990).






\end{references}
